\documentclass{aa}  
\usepackage[hyperindex=true,colorlinks=true,citecolor=blue,linkcolor=blue,breaklinks=true]{hyperref}
\usepackage{amsmath}
\usepackage{graphicx}
\usepackage{txfonts}
\usepackage{natbib,twoopt}
\usepackage{color}
\usepackage{multirow}
\usepackage{soul}
\usepackage{placeins}

\definecolor{red}{rgb}{1.,0.0,0.}

\definecolor{orange}{rgb}{1.,.65,0.}

\definecolor{vert}{rgb}{.0,.65,0.}

\bibpunct{(}{)}{;}{a}{}{,} 
\newcommandtwoopt{\citeads}[3][][]{\href{http://adsabs.harvard.edu/abs/#3}%
	{\citealp[#1][#2]{#3}}} 
\newcommandtwoopt{\citepads}[3][][]{\href{http://adsabs.harvard.edu/abs/#3}%
	{\citep[#1][#2]{#3}}} 
\newcommandtwoopt{\citetads}[3][][]{\href{http://adsabs.harvard.edu/abs/#3}%
	{\citet[#1][#2]{#3}}}
\newcommandtwoopt{\citeyearads}[3][][]%
{\href{http://adsabs.harvard.edu/abs/#3}{\citeyear[#1][#2]{#3}}}

\begin{document}
	
	\title{Probing the innermost region of the AU~Microscopii debris disc \thanks{Based on observations made with ESO telescopes at Paranal and La Silla observatory under program IDs 0101.C-0218(A), 087.C-0450(A), 087.C-0450(B) 087.C-0750(A), 088.C-0358(A) and 60.A-9165(A).}}
	\titlerunning{Probing the innermost region of the AU~Microscopii debris disc}
	
	\subtitle{}
	\author{A.~Gallenne\inst{1,2},
		C.~Desgrange\inst{3,4},
		J.~Milli\inst{3},
		J.~Sanchez-Bermudez\inst{4,5},
		G.~Chauvin\inst{2,3},
		S.~Kraus\inst{6},
		J.H.~Girard\inst{7},
		A.~Boccaletti\inst{8},
		A.M.~Lagrange\inst{8}
		\and P.~Delorme\inst{9}
	}
	
	\authorrunning{A. Gallenne et al.}
	
	\institute{
		Universidad de Concepci\'on, Departamento de Astronom\'ia, Casilla 160-C, Concepci\'on, Chile
		\and Unidad Mixta Internacional Franco-Chilena de Astronom\'ia (CNRS UMI 3386), Departamento de Astronom\'ia, Universidad de Chile, Camino El Observatorio 1515, Las Condes, Santiago, Chile
		\and Univ. Grenoble Alpes, CNRS, IPAG, F-38000 Grenoble, France
		\and Max-Planck-Institut f\"ur Astronomie, K\"onigstuhl 17, D-69117 Heidelberg, Germany
		\and Instituto de Astronom\'ia, Universidad Nacional Aut\'onoma de M\'exico, Apdo. Postal 70264, Ciudad de M\'exico 04510, Mexico
		\and School of Physics and Astronomy, University of Exeter, Exeter, Stocker Road, EX4 4QL, UK
		\and Space Telescope Science Institute, Baltimore 21218, MD, USA
		\and LESIA, Observatoire de Paris, PSL Research Univ., CNRS, Univ. Paris Diderot, Sorbonne Paris Cit\'e(c), UPMC Paris 6, Sorbonne Univ., 5 place Jules Janssen, 92195 Meudon, France
		\and Univ. Grenoble Alpes, CNRS, IPAG, F-38000 Grenoble, France
	}

	\offprints{A. Gallenne} \mail{agallenne@astro-udec.cl}
	
	\date{Received June 09, 2022; accepted June 29, 2022}
	
	
	\abstract
	{AU~Mic is a young and nearby M-dwarf star harbouring a circumstellar debris disc and one recently discovered planet on an eight-day orbit. Large-scale structures within the disc were also discovered and are moving outwards at high velocity.}
	{We aim to study this system with the highest spatial resolution in order to probe the innermost regions and to search for additional low-mass companions or set detection limits.}
	{The star was observed with two different high-angular resolution techniques probing complementary spatial scales. We obtained new $K_s$-band sparse aperture masking observations with VLT/SPHERE, which we combined with data from VLT/NACO, VLTI/PIONIER and VLTI/GRAVITY.}
	{We did not detect additional close companions within the separation range $0.02 - 7$\,au from the parent star. We determined magnitude upper limits for companions of $H \sim 9.8$\,mag within $0.02 - 0.5$\,au, $K_s \sim 11.2$\,mag within $0.4 - 2.4$\,au, and $L\arcmin \sim 10.7$\,mag within $0.7 - 7$\,au. Using theoretical isochrones, we converted these magnitudes into upper limits on the mass of $\sim17\,\mathrm{M_{jup}}$, $\sim12\,\mathrm{M_{jup}}$, and $\sim9\,\mathrm{M_{jup}}$, respectively. The PIONIER observations also allowed us to determine the angular diameter of AU~Mic, $\theta_\mathrm{LD} = 0.825 \pm 0.033_\mathrm{stat} \pm 0.038_\mathrm{sys}$\,mas, which converts to a linear radius $R = 0.862\pm0.052\,R_\odot$ when combined with the Gaia parallax.}
	{We did not detect the newly discovered planets orbiting AU~Mic ($M < 0.2\,\mathrm{M_{jup}}$), but we derived upper limit masses for the innermost region of AU~Mic. We do not have any detection with a significance beyond $3\sigma$, the most significant signal with PIONIER being $2.9\sigma$ and that with SPHERE being $1.6\sigma$. We applied the pyMESS2 code to estimate the detection probability of companions by combining radial velocities, multi-band SPHERE imaging, and our interferometric detection maps. We show that 99\,\% of the companions down to $\sim 0.5\,\mathrm{M_{jup}}$ can be detected within 0.02\,au or 1\,M$_\mathrm{jup}$ down to 0.2 au. The low-mass planets orbiting at $\lesssim 0.11$\,au ($\lesssim 11$\,mas) from the star will not be directly detectable with the current adaptive optics (AO) and interferometric instruments because of its close orbit and very high contrast ($\sim 10^\mathrm{-10}$ in $K$). It will also be below the angular resolution and contrast limit of the next Extremely Large Telescope Infrared (ELT IR) imaging instruments.}
	
	
	\maketitle
	
	%

	\section{Introduction}
	
	\object{AU~Microscopii} (AU~Mic, HD~197481, $V = 8.63$, $K_s = 4.53$, $L^\prime = 4.32$) is a young nearby M-type star, and a member of the 18.5\,Myr old $\beta$~Pictoris Moving Group \citep{Miret-Roig_2020_10_0} located at $d = 9.714\pm0.002$\,pc \citep{Gaia-Collaboration_2021_05_6,Lindegren_2021_05_5,Gaia-Collaboration_2016_11_0}. It hosts one of the best-studied edge-on debris discs and has been observed at X-ray to millimetre wavelengths \citep[see e.g.][]{Kalas_2004_03_0,Quillen_2007_10_0,Graham_2007_01_0,Schneider_2010_06_0,Wilner_2012_04_0,MacGregor_2013_01_0,Matthews_2015_10_0,Boccaletti_2015_10_0,Daley_2019_04_0,Esposito_2020_07_0} with both space- and ground-based telescopes. 
	Early modelling suggested the existence of a ring of colliding parent bodies located at  $\sim40$\,au, generating micron-sized dust particles later put on eccentric orbits through stellar wind and radiation forces, or migrating inwards through Poynting Robertson drag \citep{Augereau_2006_09_0,Strubbe_2006_09_0}. Submillimetre ALMA observations were later used to infer that this birth ring lies between $\sim8$ and 40\,au \citep{MacGregor_2013_01_0,Matthews_2015_10_0}, although the inner radius is poorly constrained. An additional halo component made of particles down to submicron sizes is seen both in scattered light and thermal imaging and extends up to $\sim 210$\,au \citep{Kalas_2004_03_0,Krist_2005_02_0,Liu_2004_09_1,Schneider_2014_10_0} with a complex morphology \citep{Fitzgerald_2007_11_0}.
	Disc features such as gaps or asymmetries created by planetesimal or infant planets make debris discs good laboratories with which to study planet formation and evolution. AU~Mic is a particularly interesting system as several morphological features are present in the disc. The disc is asymmetrical with the northwest side brighter than the southwest side, and present some substructures possibly created by larger unseen orbiting bodies \citep{Liu_2004_09_1}. Brightness enhancements have been detected and attributed to dust-density variations in the range $25-40$\,au. A prominent bump at a projected separation of $\sim 13$\,au was also identified on the southeast side above the disc plane. The disc also exhibits time variability with large-scale arc-like structures located at projected separations of $\sim 5-10$\,au on the southeast side. Those arches are moving away from the star with projected speeds that are inconsistent with a Keplerian velocity, and some even reach speeds higher than the escape velocity \citep{Boccaletti_2015_10_0,Boccaletti_2018_06_0}. One of the scenarios to explain those features would be an unseen point-source parent body releasing dust sequentially \citep{Sezestre_2017_11_0} or dust avalanches \citep{Chiang_2017_10_0}. Recent 1.3\,mm ALMA observations \citep{Daley_2019_04_0} set upper limits on the largest planetesimals stirring the disc, and ruled out a perturbing body larger than a Neptune-size object within the outer belt.
	
	Several high-contrast and high-spatial resolution observations have been reported, either using the Hubble Space Telescope (HST) or extreme adaptive optics (AO) systems, but none of these have yet been able to directly detect a planet-size body. \citet{Krist_2005_02_0} performed the first coronagraphic observations of AU~Mic but only suggested an unseen companion that would perturb the disc to explain the various detected features. Subsequent coronagraphic observations revealed more and finer details about the disc dynamic and dust distribution \citep[see e.g.][]{Wang_2015_10_0,Boccaletti_2015_10_0,Lomax_2018_02_0,Boccaletti_2018_06_0}, but still no detection of this hypothetical companion. \citet{Gauchet_2016_10_0} observed the innermost region of the AU~Mic disc in the $L^\prime$ band using the sparse aperture masking (SAM) technique, which provides a spatial resolution below the diffraction limit ($\sim \lambda/2D$). These latter authors did not find any companion brighter than $\Delta L^\prime = 5.7$\,mag ($5\sigma$ limit) within $30-500$\,mas ($0.3-5$\,au). This would correspond to a companion mass of larger than $10\,\mathrm{M_{jup}}$. From angular differential imaging with a vortex coronagraph, \citet{Launhardt_2020_03_0} reported a similar $5\sigma$ contrast limit in the $L^\prime$ band between 100\,mas and 250\,mas (6.2\,mag). Detection limits were also provided by \citet{Lannier_2017_07_0} who combined radial velocity (RV) measurements and direct imaging. These authors constrained the presence of planets with masses $\geqslant 2\,\mathrm{M_{jup}}$ from a fraction of an astronomical unit (au) of to several tens of au.
	
	Recently, a super-Neptune planet transiting AU~Mic was detected by \citet{Plavchan_2020_06_0} using NASA's Transiting Exoplanet Survey Satellite (TESS). This planet orbits the star in 8.46\,days with a semi-major axis of 0.066\,au (6.74\,mas). Such a small separation partly explains why it was not detected before with high-contrast imaging, being hidden behind the coronagraph. From RV analysis, these latter authors inferred an upper limit for the planet of $0.18\,\mathrm{M_{jup}}$ ($= 3.4\,M_\mathrm{nep}$). \citet{Klein_2021_03_0} confirmed the planet detection at a $3.9\sigma$ level with the SPIROU instrument, and inferred a mass of $0.054\,\mathrm{M_{jup}}$. \citet{Plavchan_2020_06_0} suspected a possible second transiting planet, which was later confirmed by \citet{Martioli_2021_05_8} with new TESS observations. This second planet of $\sim 0.08\,\mathrm{M_{jup}}$ orbits its star with a period of 18.9\,d and a semi-major axis of 0.11\,au ($\sim 11.3$\,mas).
	
	In this paper, we aim at setting detection limits for additional orbiting companions within the range $0.04-7$\,au ($4-730$\,mas) using high-angular-resolution observations, including infrared (IR) long-baseline interferometry and SAM. Section~\ref{section__observations_and_data_reduction} presents new SAM observations acquired in 2018 with the SPHERE/IRDIS instrument mounted at the Very Large Telescope (VLT), together with literature data from the VLTI/GRAVITY and VLT/NACO instruments. We present the details of our search for a companion and estimate our detection limits in Sect.~\ref{section__companion_search_and_detection_limits} using the \texttt{CANDID} algorithm. We then discuss our results and provide conclusions in Sect.~\ref{section__discussion_and_conclusion}.

	\section{Observations and data reduction}
	\label{section__observations_and_data_reduction}
	
	\subsection{SPHERE/SAM observations}
	
	\begin{table*}[!ht]
		\centering
		\caption{Log of our SAM observations with SPHERE/IRDIS and long-baseline interferometry with GRAVITY, together with some information about the observation of the literature data.}
		\begin{tabular}{ccccccccccc}
			\hline
			\hline
			Star			&  Date		&	MJD & DIT  & NDIT   & NEXP   &  Sp. Chan.  &  $<\mathrm{Seeing}>$	& $<\tau_0>$ & Baselines  & \# \\
			&  &	(day)	&  (s)    &          	&        &		& (\arcsec) & (ms)  &   &  \\
			\hline
			\multicolumn{10}{c}{SPHERE}  & \\
			AU~Mic 			& 2018-08-12 & 58342.22527 & 1 & 90 & 5   &  1  & 0.51 & 6.1 & -- & -- \\
			Sky					 & 2018-08-12 & 58342.23169 & 1 & 10 & 1   &  1  & 0.47 & 6.0 & -- & -- \\
			AU~Mic			& 2018-08-12 & 58342.23624 & 1 & 90 & 5   &  1  & 0.53 & 5.5 & --  & -- \\
			HD~196919	 & 2018-08-12  & 58342.24860 & 1 & 40 & 6   &  1  & 0.50 & 6.0 & -- & -- \\
			Sky					 & 2018-08-12 & 58342.25213 & 1 & 10 & 1   &  1  & 0.52 & 5.8 & --  & -- \\
			AU~Mic 			& 2018-08-12 & 58342.26256 & 1 & 30 & 15 &  1  & 0.48 & 5.8 & --  & -- \\
			Sky					 & 2018-08-12 & 58342.26847 & 1 & 10 & 1   &  1  & 0.51 & 5.1 & --   & -- \\
			AU~Mic			&  2018-08-12& 58342.27387 & 1 & 30 & 15 &  1  & 0.51 & 5.6 & --  & -- \\
			HD~196919	 & 2018-08-12 & 58342.28662 & 1 & 40 & 6   &  1  & 0.54 & 6.4 & --  & -- \\
			Sky					 & 2018-08-12 & 58342.29015 & 1 & 10 & 1   &  1  & 0.51 & 6.4 & --  & -- \\
			AU~Mic 			& 2018-08-12 & 58342.29770 & 1 & 30 & 12 &  1  & 0.47 & 6.5 & --  & -- \\
			Sky					 & 2018-08-12 & 58342.30257 & 1 & 10 & 1   &  1  & 0.41 & 7.2 & -- & -- \\
			\hline
			\multicolumn{10}{c}{NACO} & \\
			AU Mic    		& 2011-08-31 &  55804.15287 & 0.04 & 600 & 104 &  1  & 0.82  & 3.1  & -- & -- \\
			\hline
			\multicolumn{10}{c}{PIONIER}  & \\
			\multirow{22}{*}{AU~Mic}   & 2011-08-09 &  55782.30522  &  0.21  &  100  &  1  &  3  & 0.64  &  6.9  & A1-G1-I1-K0 & 1 \\
			& 2011-08-09 &  55782.31373  &  0.21  &  200  &  5  &  3  & 0.78  &  5.7  & A1-G1-I1-K0 & 1 \\
			& 2011-08-09 &  55782.32517  &  0.21  &  100  &  1  &  3  & 0.66  &  6.8  & A1-G1-I1-K0 & 1 \\
			& 2011-08-09 &  55782.34457  &  0.21  &  100  &  1  &  3  & 0.63  &  7.1  & A1-G1-I1-K0 & 1 \\
			& 2011-11-03 &  55868.01837  &  0.39  &  100  &  5  &  7  & 0.94  &  2.3  & D0-G1-H0-I1 & 2 \\                      	
			& 2011-11-03 &  55868.03855  &  0.39  &  100  &  5  &  7  & 0.64  &  3.4  & D0-G1-H0-I1 & 2 \\                      	
			& 2011-11-03 &  55868.05373  &  0.39  &  100  &  5  &  7  & 0.97  &  2.3  & D0-G1-H0-I1 & 2 \\                      	
			& 2011-11-03 &  55868.06742  &  0.39  &  100  &  5  &  7  &  1.01  &  2.2  & D0-G1-H0-I1 & 2 \\                      	
			& 2011-11-03 &  55868.07997  &  0.39  &  100  &  5  &  7  &  1.17  &  1.9  & D0-G1-H0-I1 & 2 \\                      	
			& 2011-11-03 &  55868.09390  &  0.39  &  100  &  5  &  7  & 0.81  &  2.8  & D0-G1-H0-I1 & 2 \\                      	
			& 2011-11-03 &  55868.10677  &  0.39  &  100  &  5  &  7  & 0.99  &  2.2  & D0-G1-H0-I1 & 2 \\                      	
			& 2011-11-03 &  55868.12090  &  0.39  &  100  &  5  &  7  & 0.99  &  2.3  & D0-G1-H0-I1 & 2 \\ 
			& 2013-08-09 &  56513.10329  &  0.39  &  100  &  5  &  3  & 0.92  &  2.5  & A1-B2-C1-D0 & 3 \\
			& 2013-08-09 &  56513.11264  &  0.39  &  100  &  5  &  3  & 0.76  &  3.0  & A1-B2-C1-D0 & 3 \\ 
			& 2013-08-09 &  56513.12271  &  0.39  &  100  &  5  &  3  & 0.75  &  3.0  &  A1-B2-C1-D0 & 3 \\
			& 2013-08-11 &  56515.07933  &  0.39  &  100  &  5  &  3  &  1.24  &  3.5  & A1-B2-C1-D0 & 4 \\
			& 2013-08-11 &  56515.08979  &  0.39  &  100  &  5  &  3  &  1.42  &  4.0  & A1-B2-C1-D0 & 4 \\
			& 2013-08-11 &  56515.09935  &  0.39  &  100  &  5  &  3  &  1.12  &  4.4  & A1-B2-C1-D0 & 4 \\
			& 2014-10-12 &  56942.09731  &  0.38  &  100  &  5  &  3  &  1.42  &  1.7  & A1-B2-C1-D0 & 5 \\
			& 2014-10-12 &  56942.10788  &  0.38  &  100  &  5  &  3  &  1.32  &  1.8  & A1-B2-C1-D0 & 5 \\
			& 2014-10-12 &  56942.11896  &  0.38  &  100  &  5  &  3  &  1.61  &  1.5  & A1-B2-C1-D0 & 5 \\
			& 2014-10-12 &  56942.13064  &  0.38  &  100  &  5  &  3  &  1.48  &  1.6  & A1-B2-C1-D0 & 5 \\
			\hline
			\multicolumn{10}{c}{GRAVITY} & \\	
			\multirow{3}{*}{AU~Mic}    	& 2016-06-16 &  57555.43405  &  30	  &  10	  &  5  &  1742  &  1.18  &  3.7  & A0-G1-J2-K0 &  -- \\
			& 2016-06-16  &  57555.44252  &  30	  &  10   &  5  &  1742  &  1.12  &  3.0  & A0-G1-J2-K0 & -- \\
			& 2016-06-16  &  57555.45238  &  30	  &  10   &  5  &  1742  &  1.35  &  3.0  & A0-G1-J2-K0 & -- \\
			\hline& 
		\end{tabular}
		\label{table__log}
		\tablefoot{Sp. Chan. is the number of spectral channel, $<\mathrm{Seeing}>$ is the average seeing as provided by the Differential Motion Image Monitor (DIMM) at $0.5\,\mu$m, and $<\tau_0>$ the coherence time provided by the MASS-DIMM and Baselines is the telescope configuration. The calibrators used are
			\object{HD~197540} (G8III, $V = 6.50$\,mag, $H = 4.32$\,mag, $\theta_\mathrm{LD} = 0.645\pm0.046$\,mas), 
			\object{HD~201865} (K2III, $V = 7.14$\,mag, $H = 4.04$\,mag, $\theta_\mathrm{LD} = 0.793\pm0.075$\,mas) and 
			\object{HD~202775} (K4/5III, $V = 7.49$\,mag, $H = 4.05$\,mag, $\theta_\mathrm{LD} = 0.844\pm0.072$\,mas) for \#1, 
			\object{HD~202293} (K5III, $V = 7.92$\,mag, $H = 3.74$\,mag, $\theta_\mathrm{LD} = 0.994\pm0.088$\,mas), 
			\object{HD~196387} (K4III, $V = 7.28$\,mag, $H = 3.52$\,mag, $\theta_\mathrm{LD} = 1.110\pm0.106$\,mas) and 
			\object{HD~202775} for \#2, 
			\object{HD~197540}, 
			\object{HD~194432} (K1III, $V = 7.16$\,mag, $H = 4.87$\,mag, $\theta_\mathrm{LD} = 0.555\pm0.013$\,mas) and 
			\object{HD~200750} (K1III, $V = 6.81$\,mag, $H = 4.53$\,mag, $\theta_\mathrm{LD} = 0.627\pm0.047$\,mas) for \#3,
			HD~197540, HD~194432 and HD~200750 for \#4, and 
			HD~196387 for \#5. 
		}
	\end{table*}
	
	We observed AU~Mic with the extreme AO instrument SPHERE \citep[Spectro-Polarimetric High-contrast Exoplanet REsearch,][]{Beuzit_2019_11_0} and the IRDIS/SAM mode \citep{Cheetham_2016_08_0}. This is an interferometric method \citep{Tuthill_2000_04_0} which uses a mask with non-redundant holes placed in the pupil plane. Each pair of subapertures (or baseline) will form an interference fringe pattern. Observables that are closure phases (sum of the phases between three baselines forming a triangle) and squared visibilities (fringe contrast) can then be extracted, allowing us to reach an angular resolution down to $\lambda/2D$ ($D$ is the diameter of the primary mirror).
	
	The observations were carried out on 2018 August 18 with the IRDIS \citep{Dohlen_2008_07_0} in the $Ks$-band filter ($\lambda = 2.10\,\mu$m, $\Delta\lambda = 0.30\,\mu$m) and a seven-hole aperture mask. This provides 21 baselines, that is 21 fringes with a unique spatial frequency and direction. Data were acquired in a data cube mode with tens of frames of 1\,s exposure time. Similar sequences are executed on an empty sky region for background subtraction and on a calibrator star. The full log is summarised in Table~\ref{table__log}. AU~Mic observations were interleaved with the calibrator star \object{HD~196919} (K4III, $R = 6.94$\,mag, $K = 4.31$\,mag, $\theta_\mathrm{LD} = 0.717\pm0.023$\,mas) in order to monitor the interferometric transfer function and calibrate the AU~Mic visibilities. HD~196919 was chosen with the \texttt{SearchCal} software\footnote{\url{https://www.jmmc.fr/english/tools/proposal-preparation/search-cal/}} to be as close as possible to AU~Mic in the sky and with a similar brightness.
	
\begin{figure}[h]
		\resizebox{\hsize}{!}{\includegraphics{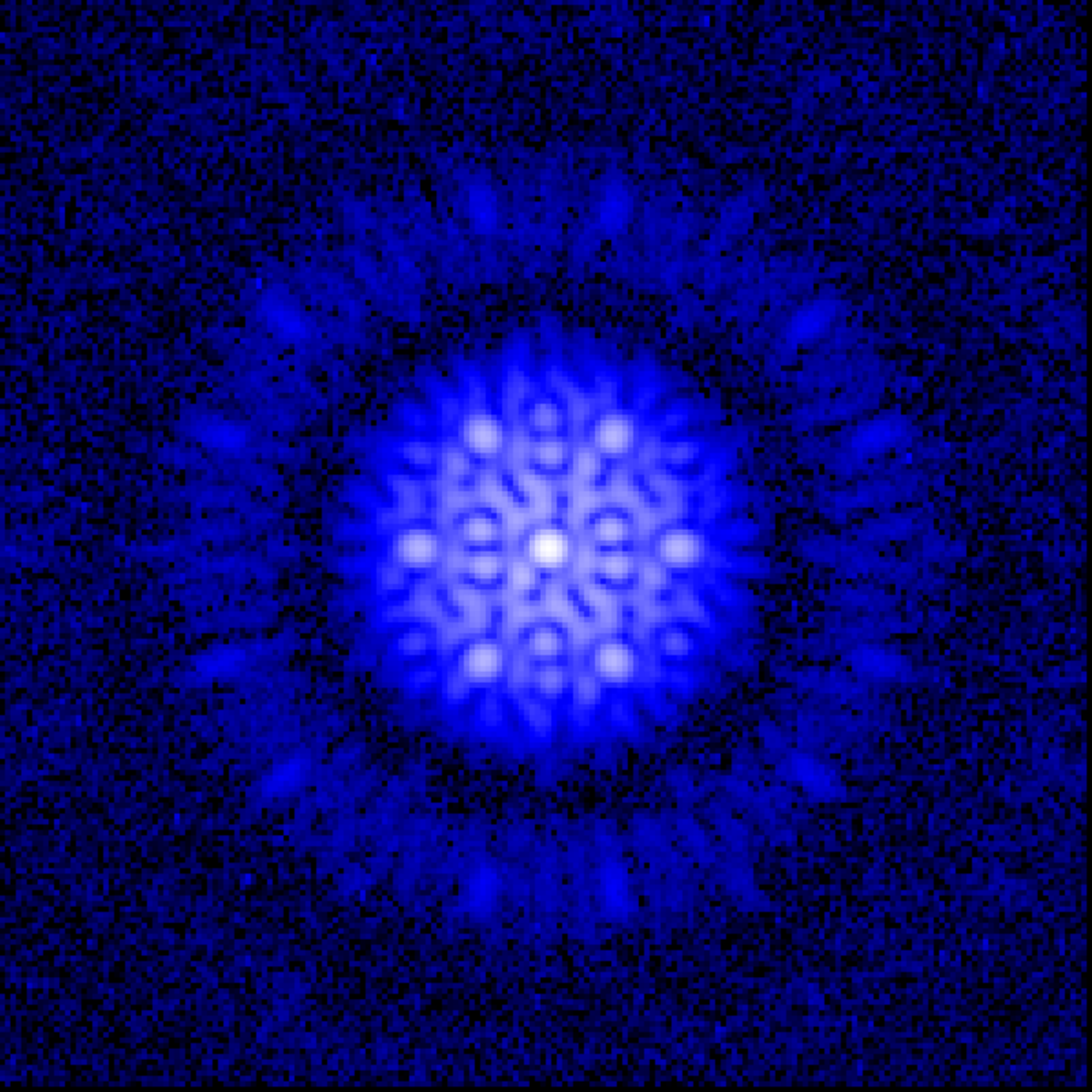}\includegraphics{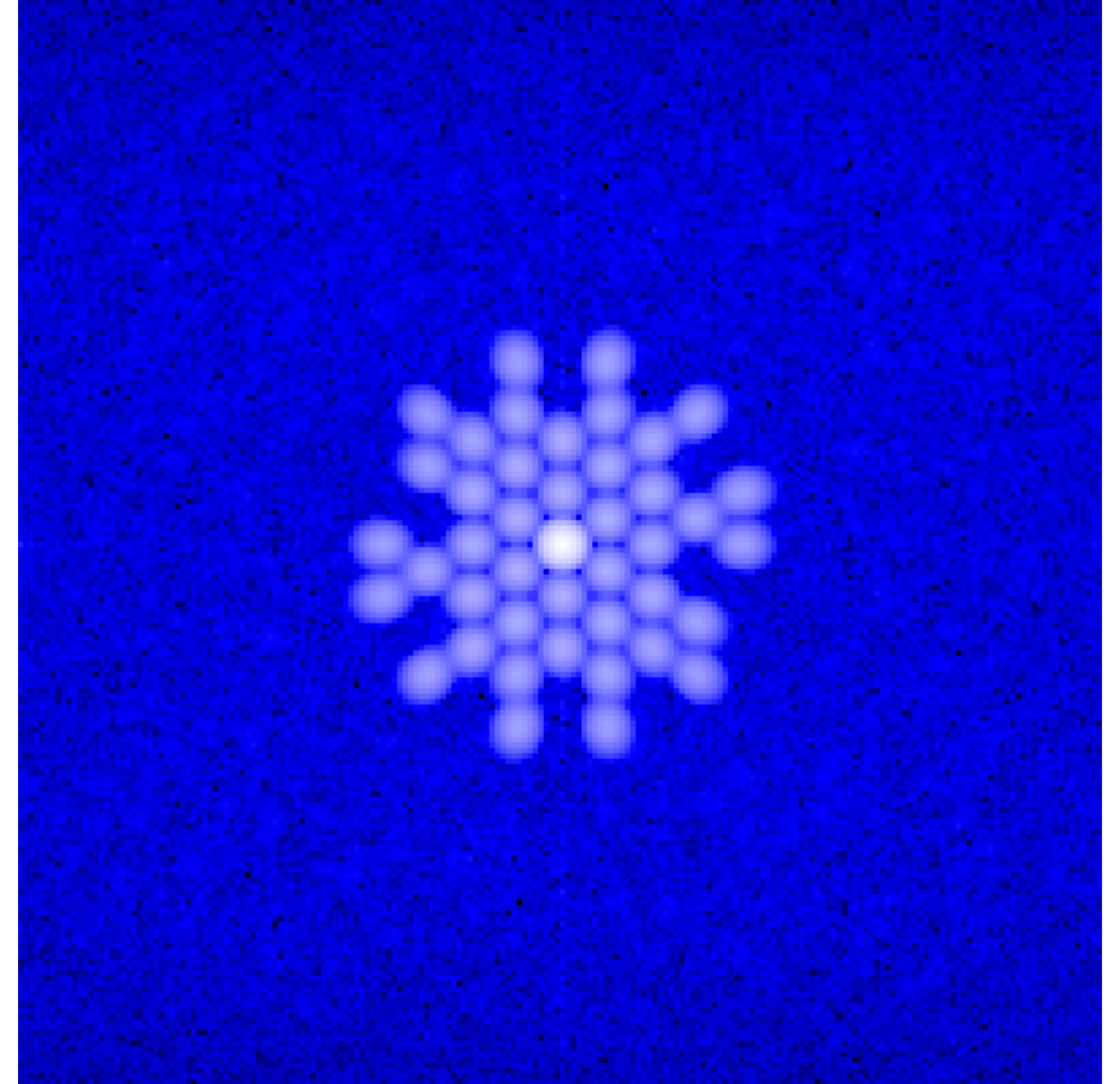}}
		\caption{Interferogram of AU~Mic trimmed within $2.45\arcsec \times 2.45\arcsec$ (left image) and its power spectrum (squared visibilities, right image).}
		\label{figure__power_spectrum}
	\end{figure}

	Our data reduction steps were as follows. First, each image was corrected for detector cosmetics using a standard procedure, including sky subtraction, flat-fielding, and bad-pixel correction. Figure~\ref{figure__power_spectrum} shows an example of the interferogram and power spectrum of AU~Mic. Subwindows of $200\times200$\,px ($2.45\arcsec \times 2.45\arcsec$) of the two images (left and right channels of the IRDIS detector) in each frame of the cubes were then extracted, stored in new cubes of 450 interferograms each, and recentred at a subpixel level. To extract complex visibilities, we used our own Python algorithm based on the formalism detailed in \citet{Tuthill_2000_04_0}. Briefly, for each cube complex visibilities were computed frame by frame from the Fourier transform of the interferograms. Squared visibilities ($V^2$) were then calculated from the squared modulus, while the bispectrum was obtained by multiplying three complex visibilities corresponding to a triangle of holes in the mask. Closure phases ($CPs$) were then obtained from the argument of the bispectrum. Calibrated $V^2$ were obtained by dividing the AU~Mic visibilities by those of the calibrator, while calibrated $CPs$ were derived by subtracting those of the calibrator. The transfer functions are displayed in Fig.~\ref{figure__tf_v2} and \ref{figure__tf_cp}. We have a total of 294 $V^2$ and 490 $CPs$ calibrated measurements, which are displayed in Fig.~\ref{figure__sphere_naco_data}. 
	
	Our SAM observations with SPHERE allow us to probe the spatial scale $\sim0.6-5.3$\,au ($60-550$\,mas) from the star, with the lower limit set by the longest baselines and the upper limit set by the bandwidth smearing. To improve the detection limits on a wider angular scale, we collected additional high-angular-resolution data from the VLTI/GRAVITY \citep{Eisenhauer_2011_03_0}, VLTI/PIONIER \citep{Le-Bouquin_2011_11_0} and VLT / NACO \citep{Rousset_2003_02_0, Lenzen_2003_03_0} instruments. 
	
	\subsection{NACO/SAM observations}
	
	We retrieved the reduced data published by \citet{Gauchet_2016_10_0}. These are also SAM mode observations and were made on 2011 August 04 with NACO in the $L^\prime$-band filter ($\lambda = 3.80\,\mu$m, $\Delta\lambda = 0.62\,\mu$m) with a seven-hole mask. The observation log from \citet{Gauchet_2016_10_0} is reported in Table~\ref{table__log}. Data reduction (flat-field, bad-pixels, and sky subtraction) were performed in the same way as for the SPHERE/SAM observations \citep[see more details in][]{Gauchet_2016_10_0}. The direct fringe-fitting method was applied instead of Fourier transform as we did previously ; however both methods provide similar results. To calibrate the data, \citet{Gauchet_2016_10_0} used \object{HD~197339} (K2III, $R = 6.47$\,mag, $K = 4.47$\,mag, $\theta_\mathrm{LD} = 0.646\pm0.021$\,mas). These data allow to probe the spatial scale $\sim0.7-7$\,au ($70-730$\,mas). Although these authors already published detection limits with this data set, we included it so that we can re-analyse the data with our \texttt{CANDID} tool, and perform a uniform analysis with the other datasets. The observations contain 546 $V^2$ and 910 $CPs$ measurements, and are displayed in Fig.~\ref{figure__sphere_naco_data}.

	\subsection{VLTI/PIONIER observations}
	
	
	PIONIER calibrated data were retrieved from the Optical interferometry Database (OiDB\footnote{\url{http://www.jmmc.fr/english/tools/data-bases/oidb/}}) developed by the Jean-Marie Mariotti Center (JMMC\footnote{\url{https://www.jmmc.fr}}). PIONIER \citep{Le-Bouquin_2011_11_0} is an interferometric instrument at the VLTI, combining the light coming from the four auxiliary telescopes (AT) in the $H$ band, either in a broad band mode or with a low spectral resolution, where the light is dispersed across a few spectral channels. 
	
	The observations were performed on several dates from 2011 to 2014 for different observing programs. Data before 2014 were collected when PIONIER was a visitor instrument at the VLTI, and therefore the calibrated data are only available in the OiDB (not in the ESO archive). We listed these observations in Table~\ref{table__log}. To monitor the instrumental and atmospheric contributions, the standard observational procedure was used which consists of interleaving the science target with reference stars. These stars are listed in Table~\ref{table__log} and taken from \texttt{SearchCal}.
	
	The data were reduced with the \texttt{pndrs} package described in \citet[][version v2.3 for 2011 data, v2.591 for 2013 and v2.71 for 2014]{Le-Bouquin_2011_11_0}. The main procedure is to compute squared visibilities and triple products for each baseline and spectral channel, and to correct for photon and readout noises. The combination of four telescopes provides six visibility and four closure phase measurements per spectral channel. Although the PIONIER observations were performed mostly with small telescope configurations, the baselines range from 10\,m to 128\,m, allowing us to probe the spatial scale $0.02 - 2$\,au ($2 - 200$\,mas ; where the interferometric FoV for our data set defined from the optical path difference separation of the fringe packet of two components). Data are displayed in Fig.~\ref{figure__pionier_gravity_data}.
	
	\subsection{VLTI/GRAVITY observations}
	
	AU~Mic was observed with GRAVITY and the Auxiliary Telescopes (ATs) on 2016 June 16 during the ESO science verification run of the instrument. GRAVITY is  a near-IR interferometric beam combiner operating in the $K$-band ($\lambda = 1.990-2.450\,\mu$m). The observations were carried out using the highest spectral resolution ($R \sim 4000$) and the telescope baselines A0-G1-J2-K0 (projected baselines range from 48\,m to 110\,m). Three datasets of 300\,s each ($10\times30$\,s) interleaved by sky exposures were recorded. To monitor the interferometric transfer function, the calibrator \object{HD~155276} (K0.5IIICN0.5, $K = 3.93$\,mag, $\theta_\mathrm{LD} = 0.799\pm0.078$\,mas) was observed (taken from \texttt{SearchCal}.). However, HD~155276 was observed about 6\,h before AU~Mic, and therefore the calibration of the squared visibilities is likely unreliable. However, the closure phases can be used as they are independent of telescope-specific phase shifts induced by the atmosphere or optics \citep[see e.g.][]{Monnier_2007_10_6}. Observables (visibilities and closure phases) were extracted using the provided GRAVITY pipeline \citep[v1.1.2,][]{Lapeyrere_2014_07_0}. The projected baselines allow us to probe the inner region of $0.03-2.4$\,au ($3.5-250$\,mas) from the central star (FoV limited by the fiber FoV and not the bandwidth smearing thanks to the high spectral resolution). Data are displayed in Fig.~\ref{figure__pionier_gravity_data}.
	

	\section{Companion search and detection limits}
	\label{section__companion_search_and_detection_limits}
	
	\subsection{The CANDID algorithm}
	
	To search for a possible companion and set detection limits, we used the interferometric tool \texttt{CANDID}\footnote{Available at \url{https://github.com/amerand/CANDID} and \url{https://github.com/agallenne/GUIcandid} for a GUI version.} \citep{Gallenne_2015_07_0}. The main function allows a systematic search for companions (point-sources in this case) performing an $N\times N$ grid of fits, the minimum required grid resolution of which is estimated a posteriori in order to find the global minimum in $\chi^2$. The tool delivers the binary parameters, namely the flux ratio $f$, and the relative astrometric separation $(\Delta \alpha, \Delta \delta)$. The uniform disc (UD) angular diameters of both components can also be fitted, but in our case we do not expect a spatially resolved companion. Furthermore, the angular diameter $\theta_\mathrm{UD}$ of AU~Mic is only resolved by GRAVITY and PIONIER. The significance of the detection is also given, taking into account the reduced $\chi^2$ and the number of degrees of freedom.
	
	The second main function of \texttt{CANDID} incorporates a robust method to set a $n\sigma$ detection limit on the flux ratio for undetected components, which is based on an analytical injection of a fake companion at each point in the grid. We refer the reader to \citet{Gallenne_2015_07_0} for more details about \texttt{CANDID} and its formalism. Two-dimensional detection-limit maps are expressed in contrast ratio in the corresponding bandpass and spatial range of the instrument. To have a quantitative estimate of the sensitivity limit with respect to the separation, we estimated a radial profile using the 90\,\% completeness level (i.e. 90\,\% of all possible positions) from the cumulated histogram in rings for all azimuths.
	
	\subsection{Angular diameter of AU~Mic}
	
	To measure the diameter, we only used the \#1 PIONIER observations because this is the only dataset with long enough baselines to allow a reliable measurement. In addition, PIONIER operates in the $H$ band and therefore provides a better spatial resolution than GRAVITY. There is no bright companion orbiting AU~Mic reported, and therefore we do not expect the angular diameter measurement to be biased by the presence of a low-contrast component.
	
	We first estimated the uniform disc diameter $\theta_\mathrm{UD}$ by fitting the calibrated squared visibilities with the following formula:
	\begin{equation}
		|V(u,v)|^2 = \left| \dfrac{2~J_1(x)}{x} \right|^2
	\end{equation}
	where $J_1(x)$ is the first-order Bessel function, $x = \pi \theta_\mathrm{UD} B/\lambda$, $\lambda$ the wavelength, $B = \sqrt{u^2 + v^2}$ the baseline length and $(u,v)$ are the spatial frequencies. We measured $\theta_\mathrm{UD} = 0.797 \pm 0.032_\mathrm{stat} \pm 0.038_\mathrm{sys}$\,mas. The statistical error was determined using the bootstrapping technique (with replacement) on all baselines. The given diameter corresponds to the median of the distribution, while the maximum value between the 16th and 84th percentile was chosen as the statistical uncertainty. The systematic error is the quadratic sum of two systematic uncertainties. The first comes from the calibrator uncertainties ; we took the quadratic average of the calibrator uncertainties used. The second is a 0.35\,\% systematic error due to the uncertainty of the wavelength calibration of PIONIER \citep{Gallenne_2012_03_0}. We used this value in the following \texttt{CANDID} analysis. Our measurement is in good agreement with the $0.78\pm0.023$\,mas reported by \citet[][we assume they report the uniform disc diameter as there is no reference to any limb-darkening calculation]{White_2015_01_0}.
	
	We additionally tested a UD model including an IR emission from the hot dust in the inner disc. Such emission is usually difficult to detect because the flux ratio $f_\mathrm{d} = f_\mathrm{disc}/f_\star$ is of the order of 1\,\% which needs high-precision visibilities and small baselines. We fitted a simplified visibility model $(1 - 2f_\mathrm{d})*|V(u,v)|^2$ which assumes the disc is fully resolved resulting in a visibility deficit compared to the values expected from the star alone. We found $\theta_\mathrm{UD} = 0.679\pm0.12$\,mas and $f_\mathrm{d} = 1.09\pm0.67$\,\%. The angular diameter is smaller but within $1\sigma$ with the single star value. In addition, $f_\mathrm{d}$ is barely constrained with a $\chi^2_r$ difference of 0.2 only, corresponding to a significance level of $1.8\sigma$. We can therefore assume that the single star model provides an unbiased angular diameter.
	
	We determine the limb-darkened angular diameters, $\theta_\mathrm{LD}$. Assuming a circular symmetry, we fitted the calibrated squared visibilities following the formalism of \citet{Merand_2015_12_0}, which consists in extracting the radial intensity profile $I(r)$ of the spherical \texttt{SATLAS} models \citep{Neilson_2013_06_0}, which was converted to a visibility profile using a Hankel transform:
	\begin{equation}
		V_\lambda (x) = \frac{\int_0^1 I_\lambda(r) J_0(rx) r dr}{ \int_0^1 I_\lambda(r) r dr },
	\end{equation}
	where $\lambda$ is the wavelength, $x = \pi \theta_\mathrm{LD} B/\lambda$, $B$ is the interferometric baseline projected onto the sky, $J_0$ the Bessel function of the first kind, and $r = \sqrt{1-\mu^2}$, with $\mu = \cos(\theta)$, $\theta$ being the angle between the line of sight and a surface element of the star.
	
\begin{figure}[h]
		\resizebox{\hsize}{!}{\includegraphics{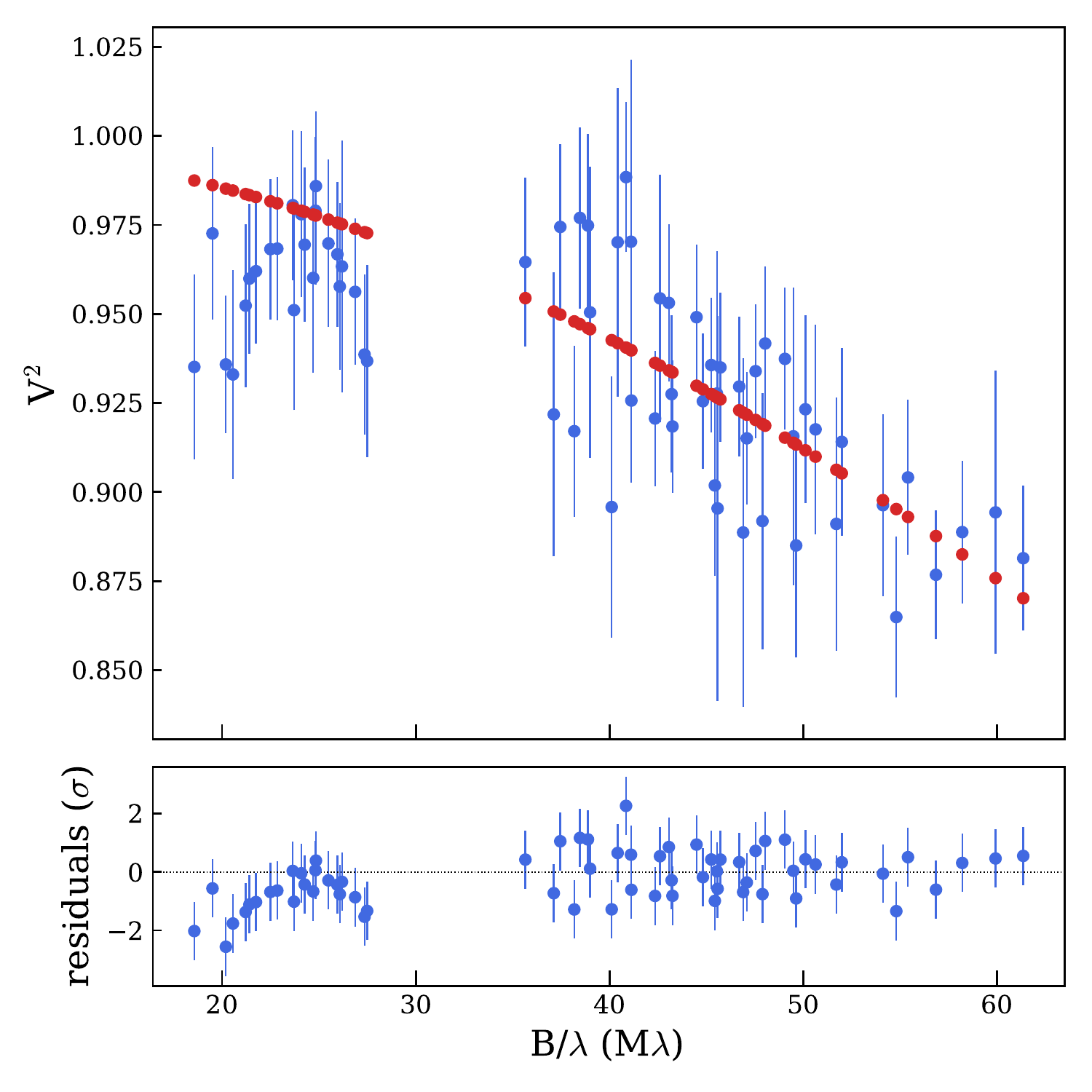}}
		\caption{Best-fit solution of the limb-darkened disc model (red dots) to the PIONIER squared visibilities of AU~Mic (blue dots).}
		\label{figure__angular_diameter}
	\end{figure}

	The \texttt{SATLAS} grid models span effective temperatures from 3000 to 8000\,K in steps of 100\,K, effective gravities from -1 to 3 in steps of 0.25, and masses from 0.5 to 20\,$M_\odot$. We chose the model with the closest parameters, that is with $T_\mathrm{eff} = 3700$\,K, $\log g = 4.5$, and $M  = 0.5\,M_\odot$ \citep[][with $\log{g} \sim 4.4$\,dex estimated from the mass, temperature and luminosity]{Plavchan_2020_06_0}. We estimated $\theta_\mathrm{LD} = 0.825 \pm 0.033_\mathrm{stat} \pm 0.038_\mathrm{sys}$\,mas and errors were estimated in the same way as above. Changing the models with $T_\mathrm{eff} \pm 200$\,K, $\log g\pm 0.5$, and $M \pm 2.5\,M_\odot$ changes the diameters by at most 0.1\,\%, which we also quadratically added. Combining our estimate with the Gaia parallax of $102.943\pm0.023$\,mas provides a linear radius of $R = 0.862\pm0.052\,R_\odot$. The best-fit solution for the angular diameter is shown in Fig.~\ref{figure__angular_diameter}.
	
	We also checked the angular diameter using the squared visibilities from GRAVITY. We found $\theta_\mathrm{LD} = 1.078 \pm 0.255_\mathrm{stat} \pm 0.078_\mathrm{sys}$\,mas (systematic error from the uncertainty of the calibrator). This value is higher than the one determined with PIONIER, although within $1\sigma$. The squared visibilities are not very accurate or precise, with a standard deviation of $\sim 10$\,\% (see Fig.~\ref{figure__pionier_gravity_data}). This is likely linked to a poor calibration, which was somewhat expected as the calibrator was observed 6\,h before. However, as stated previously, the closure phases are independent of atmosphere seeing, and therefore we use only this observable for the following analyses.
	
	\subsection{Search for orbiting companion}
	
	\begin{figure*}[h]
		\resizebox{\hsize}{!}{\includegraphics{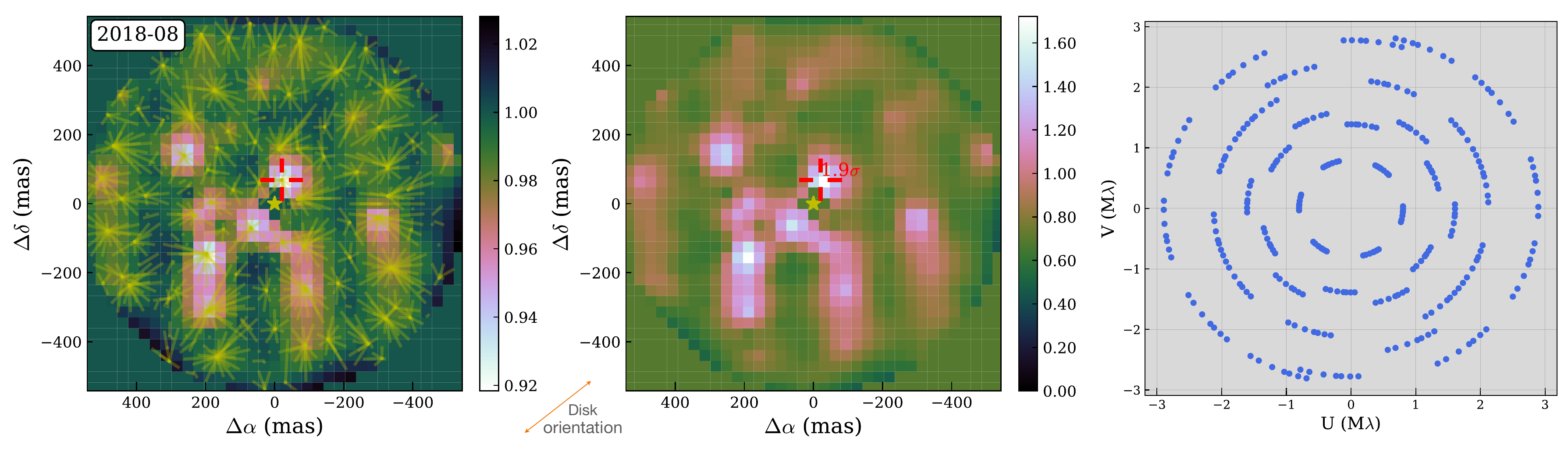}}
		\caption{$\chi^2_r$ map of the local minima for the SPHERE observations (left panel), detection level map (middle panel), and the $(u,v)$ plane coverage of the observations (right panel). The yellow lines represent the convergence from the starting points to the final fitted position. The maps were re-interpolated in a regular grid for clarity. The yellow star denotes AU~Mic, while the orange arrow in the middle indicates the disc orientation.}
		\label{figure__sphereChi2Map}
	\end{figure*}	
	
	Every instrument has a different minimum and maximum spatial range $r$ within which we can search for a companion. The loss of coherence caused by spectral smearing mainly limits the maximum separation for a reliable detection ($\lesssim R\lambda/B$), while the smallest spatial scale is set by the angular resolution ($\sim\lambda/2B$). The ranges are mentioned in Sect.~\ref{section__observations_and_data_reduction} ; we determined $r \sim 60 - 550$\,mas for SPHERE, $r \sim 70 - 730$\,mas for NACO, $r \sim 2 - 200$\,mas for PIONIER and $r \sim 3.5 - 250$\,mas for GRAVITY. In the following, we searched for companions using $CPs$ only. As explained by \citet{Gallenne_2015_07_0}, the $CPs$ are more sensitive to faint off-axis companions (although depending on its location and the $(u,v)$ coverage), and are also less affected by instrumental and atmospheric perturbations. The angular diameter is fixed in the fitting process to the uniform disc value listed above.
	
	In the following, the contrasts have been converted to masses using the AMES-COND models \citep{Baraffe_2003_05_0}.
	
	\paragraph{SPHERE data:} We did not detect any companion at more than $1.9\sigma$. Figure~\ref{figure__sphereChi2Map} shows the 2D detection maps given by \texttt{CANDID}. The highest detection is at $\rho \sim 72$\,mas and $PA \sim -17^\circ$ with a flux ratio $f_\mathrm{K} = 0.041$\,\%, which would correspond to a companion mass of $\sim7.3\,M_\mathrm{j}$. However, even if the position seems slightly above the disc plane \citep[$PA = 128.41^\circ$,][]{MacGregor_2013_01_0}, this detection is not significant. The $(u,v)$ coverage, that is, the sampling of the Fourier plane, is displayed in Fig.~\ref{figure__sphereChi2Map}.
	
	\paragraph{NACO data:} A clear location at $\rho \sim 139$\,mas and $PA \sim -156^\circ$ is detected in the $\chi^2$ map but the detection level is only $1.6\sigma$ (see maps in Appendix~\ref{appendix__detection_maps}). The flux ratio is $f_\mathrm{L} = 0.07$\,\%  ($\sim 4.2\,M_\mathrm{j}$) but its location perpendicular to the disc plane makes this detection unlikely, although similar in brightness with the previous highest detection of SPHERE. 
	
	\paragraph{PIONIER data:} We analysed all dates independently. There is also no detection with a statistical significance larger than $2.9\sigma$ (see maps in Appendix~\ref{appendix__detection_maps}). The highest peak has a flux ratio of $f_\mathrm{H} = 0.49$\,\%  ($\sim 14.1\,M_\mathrm{j}$) and is located at $\rho \sim 44$\,mas and $PA \sim -109^\circ$. We can see additional peaks but these are spuriously produced by an incomplete $(u, v)$ coverage, and their significance is less than $2.9\sigma$.
	
	\paragraph{GRAVITY data:} Only $CPs$ from the science combiner have been used as it offers a much greater number of spectral channels than the fringe tracker (five channels only). No additional sources are detected at more than $2.6\sigma$. The highest peak is at $\rho \sim 14.8$\,mas, $PA \sim -128^\circ$, and with a flux ratio of $f_\mathrm{K} = 0.42$\,\%  ($\sim 13.5\,M_\mathrm{j}$). This companion is unlikely as it is too bright, and the flux ratio is not consistent with the SPHERE estimate.

	We therefore did not significantly detect (i.e. $> 5\sigma$) any component orbiting AU~Mic. The highest probable locations given by the PIONIER and GRAVITY data are unlikely as they would correspond to a brown dwarf. Although such objects would not have been detected from AO imaging because of the limited angular resolution and size of the coronagraph, it would have been detected from the extensive observing RV observations (from Kepler's law and assuming our measured projected separation as a lower limit for the angular semi-major axis, the minimum RV amplitude would be $\sim 0.8\,\mathrm{km~s^{-1}}$).

	\subsection{$5\sigma$ detection limits}
	
\begin{figure}[h]
		\resizebox{\hsize}{!}{\includegraphics{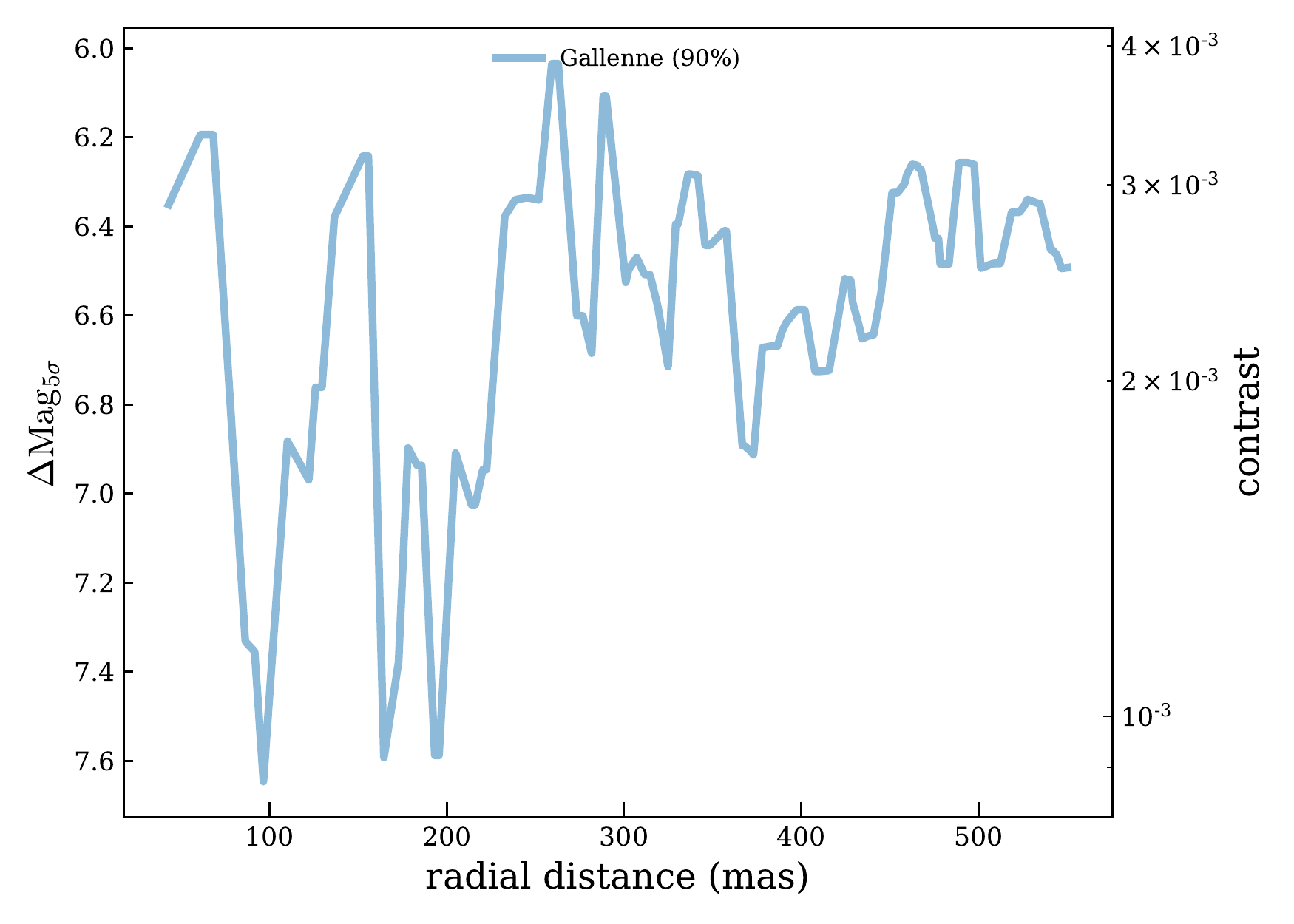}}
		\caption{Contrast limit at $5\sigma$ for our SPHERE observations.}
		\label{figure__detection_limit}
	\end{figure}

	\texttt{CANDID} has also implemented a robust method to derive the dynamic range we can reach with a given set of data. It consists in injecting a fake companion into the data at each astrometric position with different flux ratios. As we inject a companion, we therefore know that the binary model should be the true model. We then compare the $\chi^2$ with that of a single star model (uniform disc model) to obtain the probability of the binary model being the true model. We set the significance level on the flux ratios at $5\sigma$, meaning that lower flux ratios are not significantly detected. Doing this for all points in the grid, we then have a $5\sigma$ detection limit map for the flux ratio. We refer the reader to \citet{Gallenne_2015_07_0} for more information about the method. To obtain a quantitative estimate of the sensitivity limit with respect to the separation, we estimated a radial profile using the 90\,\% completeness level (i.e. 90\,\% of all possible positions) from the cumulated histogram in rings for all azimuths. The radial profile for the detection limit of our SPHERE observations is displayed in Fig.~\ref{figure__detection_limit}.
	
	We list three different values in Table~\ref{table__detection_limits}: the average limit for the separation ranges $r < r_\mathrm{mid}$, $ r_\mathrm{mid} < r < r_\mathrm{max}$, and the full range $ r < r_\mathrm{max}$. All of the final $5\sigma$ contrast limits, $\Delta m_{5\sigma}$ expressed in magnitude, are conservative as they correspond to the mean plus the standard deviation for the given radius range. Upper limit masses, that is, the lowest masses we can safely exclude, were estimated from the contrast limits calculated for the full separation range.
	
	\begin{table*}[!ht]
		\centering
		\caption{Average $5\sigma$ contrast limits ($\Delta m_\lambda$).}
		\begin{tabular}{cccccc}
			\hline
			\hline
			&  Date	  &		\multicolumn{3}{c}{$\Delta m$}  & upper limit mass		   \\
			&			 &		\multicolumn{3}{c}{(mag)} &    $\mathrm{M_{jup}}$ 	  \\
			\hline
			&            & $0.4 < r < 1$\,au  & $1 < r < 2.4$\,au & $0.4 < r < 2.4$\,au &  \\   
			SPHERE ($K_\mathrm{s}$)  & 2018-08  & 6.67  & 6.72 & 6.68 & 12  \\
			\hline
			&            & $0.7 < r < 3.9$\,au  & $3.9 < r < 7$\,au & $0.7 < r < 7$\,au &  \\   
			NACO ($L^\prime$)  & 2011-08  & 6.51  & 6.32  & 6.33 & 9 \\
			\hline
			&            & $0.02 < r < 0.25$\,au  & $0.25 < r < 0.5$\,au & $0.02 < r < 0.5$\,au &  \\   
			PIONIER ($H$)			  	& 2011-11  		& 4.87  & 5.10 & 5.02 & 17 \\
			\hline
			&            & $0.03 < r < 1.4$\,au  & $1.2 < r < 2.4$\,au & $0.03 < r < 2.4$\,au &  \\   
			GRAVITY ($K$)  & 2016-06  &  5.25 & 5.52  & 5.39 & 16 \\
			\hline
		\end{tabular}
		\label{table__detection_limits}
	\end{table*}
	
	\section{Discussion and Conclusion}
	\label{section__discussion_and_conclusion}
	
	With our new SPHERE observations, we do not detect additional objects brighter than $K_s \sim 11.2$\,mag within $0.4 - 2.4$\,au, which would correspond to a mass limit for a planet of $12.3\pm0.5\,\mathrm{M_{jup}}$. The mass limit was estimated using a cubic spline interpolation of the AMES-COND isochrones encompassing 18.5\,Ma (i.e. 9, 10, 20, and 30\,Ma). We then interpolated the mass for the corresponding magnitude limit and age of $18.5\pm2$\,Ma. The best mass sensitivity is achieved with NACO because the $L^\prime$ band is favourable at young ages, providing an upper limit on the mass of $\sim 8.3\pm0.8\,\mathrm{M_{jup}}$ within $0.7 - 7.3$\,au. This is similar to the upper mass limit of $10.1\,\mathrm{M_{jup}}$ determined by \citet{Gauchet_2016_10_0} with the same data set, but updated with our more robust \texttt{CANDID} tool. The SAM observations enable us to reach higher contrast thanks to a better AO correction and a more stable point spread function. In addition, the light beam in SPHERE is prone to a higher throughput than the whole VLTI optical path. Long-baseline interferometric observations enable us to probe smaller spatial scales but the contrast limit reached is not as good. Within $0.02 - 0.5$\,au, PIONIER provides a detectable mass limit of $\sim 17.1\pm1.0\,\mathrm{M_{jup}}$.
	
\begin{figure*}[h]
		\resizebox{\hsize}{!}{\includegraphics{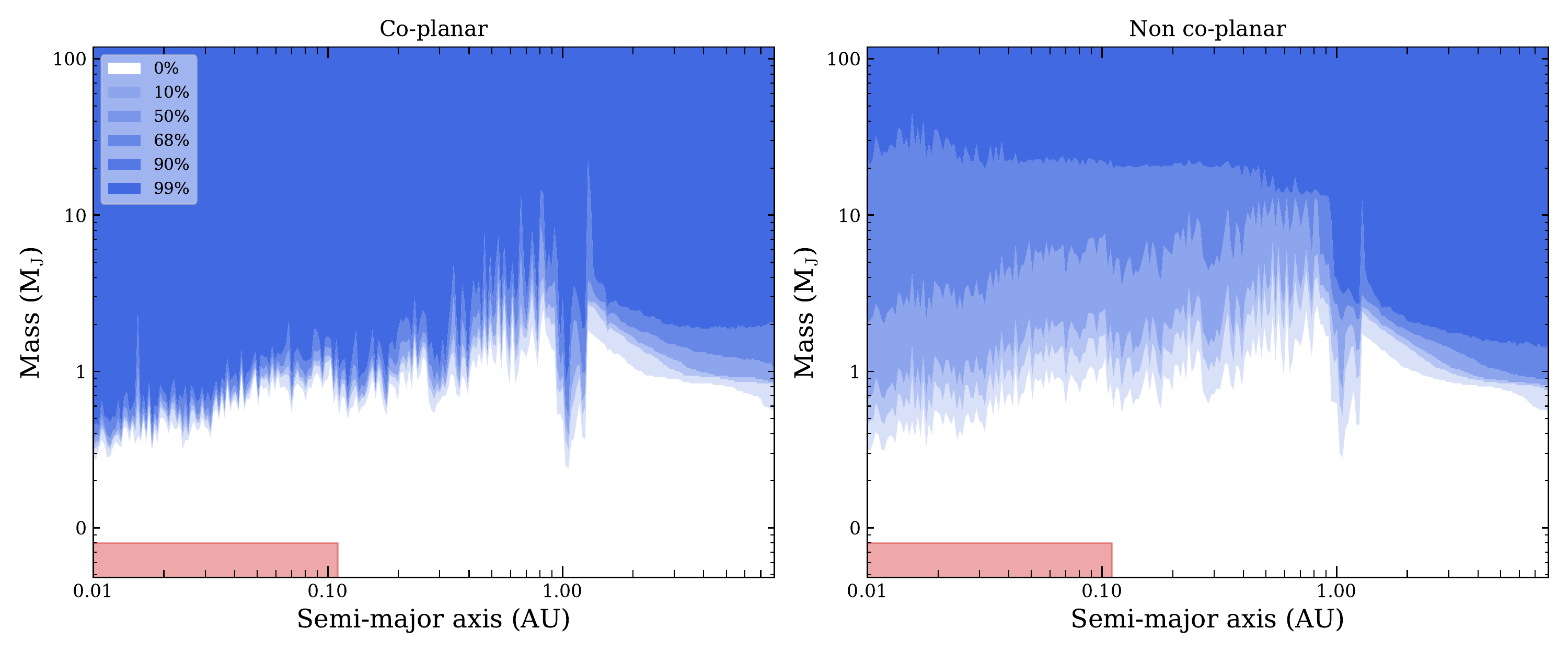}}
		\caption{Detection probability maps of AU~Mic assuming coplanarity with the disc (left panel) or with no hypothesis on the inclination (right panel), combining RV, interferometric, and direct-imaging observations. Blue colour-coded areas denote the detectability probability. The red area shows where the newly detected companions should be located \citep{Martioli_2021_05_8}.}
		\label{figure_mess2map}
	\end{figure*}

\begin{figure*}[h]
		\resizebox{\hsize}{!}{\includegraphics{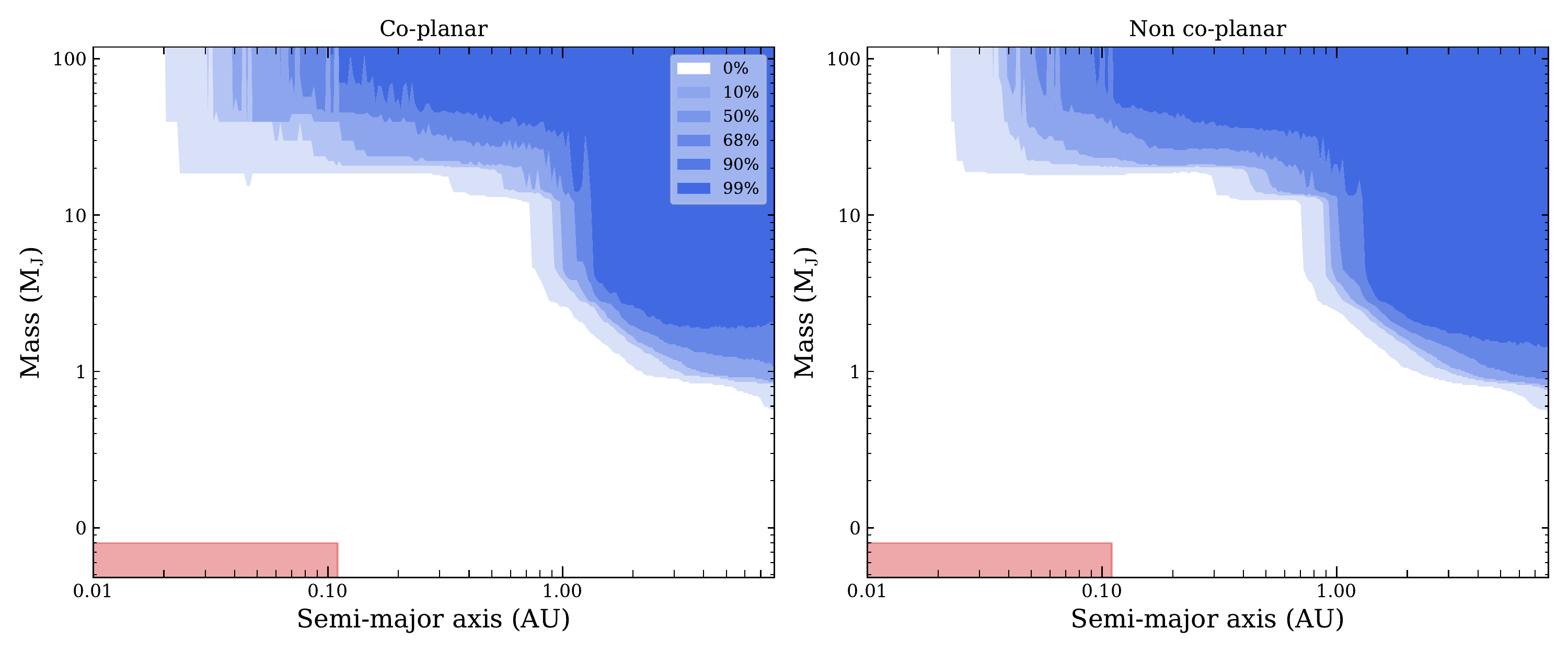}}
		\caption{Same as Fig.~\ref{figure_mess2map} but only interferometric and direct-imaging observations.}
		\label{figure_mess2map2}
	\end{figure*}

	We added additional constraints on the mass limits by combining our calculated $5\sigma$ detection maps with the existing RV data from \citet{Lannier_2017_07_0} and SPHERE contrast maps estimated from the data of \citet[][determined using a PCA analysis of the IFS and IRDIS images taken in the $J, H, H23$, and $K12$ bands]{Boccaletti_2018_06_0}.  We used the Multi-epoch multi-purpose Exoplanet Simulation System algorithm \citep[MESS2,][]{Lannier_2017_07_0} which generates populations of synthetic planets through a Monte Carlo simulation for a given range of mass and separation. MESS2 grids are then compared to our combined data sets to explore the detection probability. Our contrast limit maps were converted into masses using an age of 18.5\,Ma \citep{Miret-Roig_2020_10_0} and the COND-2003 evolutionary models \citep{Baraffe_2003_05_0}. We considered two inclination distributions for the synthetic planets: a uniform distribution between 0$^\circ$ and 90$^\circ$, and an inclination of 90$\pm$1$^\circ$, that is, assuming the synthetic planets orbit in the same plane as the resolved disc. We also took an ad hoc eccentricity distribution with $e < 0.6$. The probability maps are displayed in Fig.~\ref{figure_mess2map}.  As expected, assuming a uniform prior on the inclination significantly reduces the sensitivity to companions. This is particularly true within 1\,au where the sensitivity is driven by the RV technique, which is more sensitive to edge-on systems. In the case of co-planar orbits,  99\,\% of additional companions with $\gtrsim 4$\,M$\mathrm{_{jup}}$ can be excluded within $0.2 - 1$\,au, and $\gtrsim 2$\,M$\mathrm{_{jup}}$ within $1 - 8$\,au. These detection limits are not better than the ones derived by \citet{Lannier_2017_07_0} combining RVs and direct imaging at separation $r > 0.2$\,au. This is because direct imaging has a better sensitivity than interferometry at these angular separations. However, for the innermost region, at $\lesssim 0.2$\,au, which is the spatial scale probed by interferometry, we find that 99\,\% of the companions down to $\sim 0.5$\,M$\mathrm{_{jup}}$ can be detected. This is actually similar to using RVs alone because of the lack of high-dynamic range of the current interferometric instruments and our limited datasets. In this specific case (i.e. for co-planar orbits of the edge-on disc), interferometry does not bring additional constraints when precise RVs are available. In a more general case where no precise RVs are available, interferometry is complementary to direct imaging by providing better contrast in the innermost region, as we can see in Fig.~\ref{figure_mess2map2}. We also notice that the coplanarity hypothesis mostly affects the RVs, and has a limited effect when only interferometry and direct imaging are used.
	
	
	The newly detected planets have a mass of $< 0.08\,\mathrm{M_{jup}}$ for a maximum separation of 11.3\,mas (0.11\,au), well below our contrast limits (red area in Fig.~\ref{figure_mess2map}). This would correspond to a lower limit $K$-band magnitude of $\sim 33$\,mag, corresponding to a contrast of $\sim 4\times10^{-12}$ ($\Delta K \sim 28.5$\,mag, extrapolated from the AMES-COND models). Such dynamic range is beyond the capabilities of the current imaging and interferometric instruments, and also beyond future facilities in the near future. Only space-based coronagraph like Habex or Louvoir \citep[see e.g.][]{Mennesson_2016_07_0,Bolcar_2017_09_0} could reach such contrasts but they would not have the required angular resolution. Direct imaging of these planets with AO is not possible due to its proximity with the host star. Only long-baseline interferometry can probe the innermost regions, but the current reachable contrast is about $\Delta H \sim 6.5$\,mag \citep{Gallenne_2015_07_0,Roettenbacher_2015_08_0}. The dual-field mode of GRAVITY reached a contrast of $\Delta K = 10.7$\,mag by directly measuring a spectrum of an exoplanet \citep{Gravity-Collaboration_2019_03_0}, but this is a special case as the location of the companion must be known a priory to be able to position the fibre on it. A 'blind search' was tested on AU~Mic with the Unit Telescopes by dithering the fibre along the southeastern part of the debris disc \citep{GRAVITY-Collaboration_2019_12_0}. No planet was detected but they achieved a $5\sigma$ dynamic range of 11\,mag at 1.2\,au and 13.5\,mag at 2.4\,au. The astrometric mode of GRAVITY, which can provide a differential astrometric precision of a few tens of $\,\mu$as, could also be an alternative used to obtain the astrometric orbit of the more massive planet ; however the astrometric wobble produced by this planet on the central star is too small ($\sim 1.7\,\mu$as, taking the upper limit of $0.08\,\mathrm{M_{jup}}$ at 0.11\,au).
	
	The upcoming ELT with a SAM mode will likely also not detect these planets. Assuming a maximum baseline of 38\,m (for the possibly two most distant mask holes) and observations performed in the $K$ band, planets detectable down to $\sim 7$\,mas would need an optimal AO correction and strongly improved sensitivity limits (although near the SAM resolution limit). Unfortunately, even the next generation of extreme-AO instruments for the ELT will not provide a contrast of better than $10^{-7}$ in such close proximity to the star.

	
	\begin{acknowledgements}
		A.G. acknowledges the support of the French Agence Nationale de la Recherche (ANR), under grant ANR-15-CE31-0012-01 (project UnlockCepheids). A.G. acknowledges financial support from "Programme National de Physique Stellaire" (PNPS) of CNRS/INSU, France. A.G. also acknowledges financial support from the ANID-ALMA fund No. ASTRO20-0059. The research leading to these results has received funding from the European Research Council (ERC) under the European Union's Horizon 2020 research and innovation programme under grant agreement No 695099 (project CepBin), grant agreement No 951549 (project UniverScale) and from the National Science Center, Poland grants MAESTRO UMO-2017/26/A/ST9/00446 and BEETHOVEN UMO-2018/31/G/ST9/03050. A.G. acknowledges support from the IdP II 2015 0002 64 and DIR/WK/2018/09 grants of the Polish Ministry of Science and Higher Education, from the BASAL Centro de Astrof\'isica y Tecnolog\'ias Afines (CATA) AFB-170002, and from the Millennium Institute of Astrophysics (MAS) of the Iniciativa Cient\'ifica Milenio del Ministerio de Econom\'ia, Fomento y Turismo de Chile, project IC120009. S.K. acknowledges support from an ERC Consolidator Grant (Grant Agreement ID 101003096). The GRAVITY data have been recorded as part of the Science Verification program and we thank the SV team, which is composed of ESO employees and GRAVITY consortium members, for carrying out the observations (\url{https://www.eso.org/sci/activities/vltsv/gravitysv.html}). This work has made use of data from the European Space Agency (ESA) mission {\it Gaia} (\url{https://www.cosmos.esa.int/gaia}), processed by the {\it Gaia} Data Processing and Analysis Consortium (DPAC, \url{https://www.cosmos.esa.int/web/gaia/dpac/consortium}). Funding for the DPAC has been provided by national institutions, in particular the institutions participating in the {\it Gaia} Multilateral Agreement.
	\end{acknowledgements}
	
	
	\bibliographystyle{aa}   
	\bibliography{/Users/alex/Sciences/Articles/bibliographie}
	
	
	\begin{appendix}
	
		\section{Transfer functions of our SPHERE/SAM observations}
		
		\begin{figure*}[h]
			\resizebox{\hsize}{!}{\includegraphics{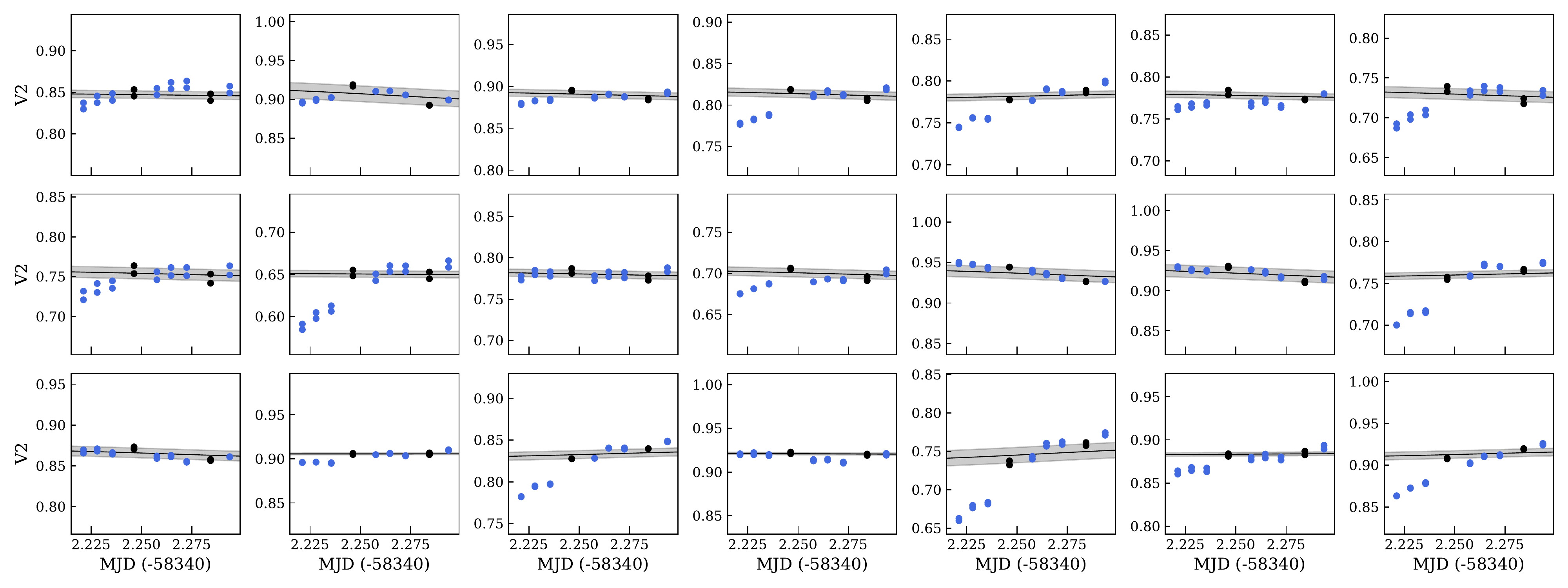}}
			\caption{Transfer function calibration for the squared visibilities. Calibrator and science target observations are represented with black and blue dots, respectively. The solid line denotes the interpolation of the transfer function, while the grey area represents the standard deviation with the calibrators.}
			\label{figure__tf_v2}
		\end{figure*}
		\FloatBarrier
		\begin{figure*}[h]
			\resizebox{\hsize}{!}{\includegraphics{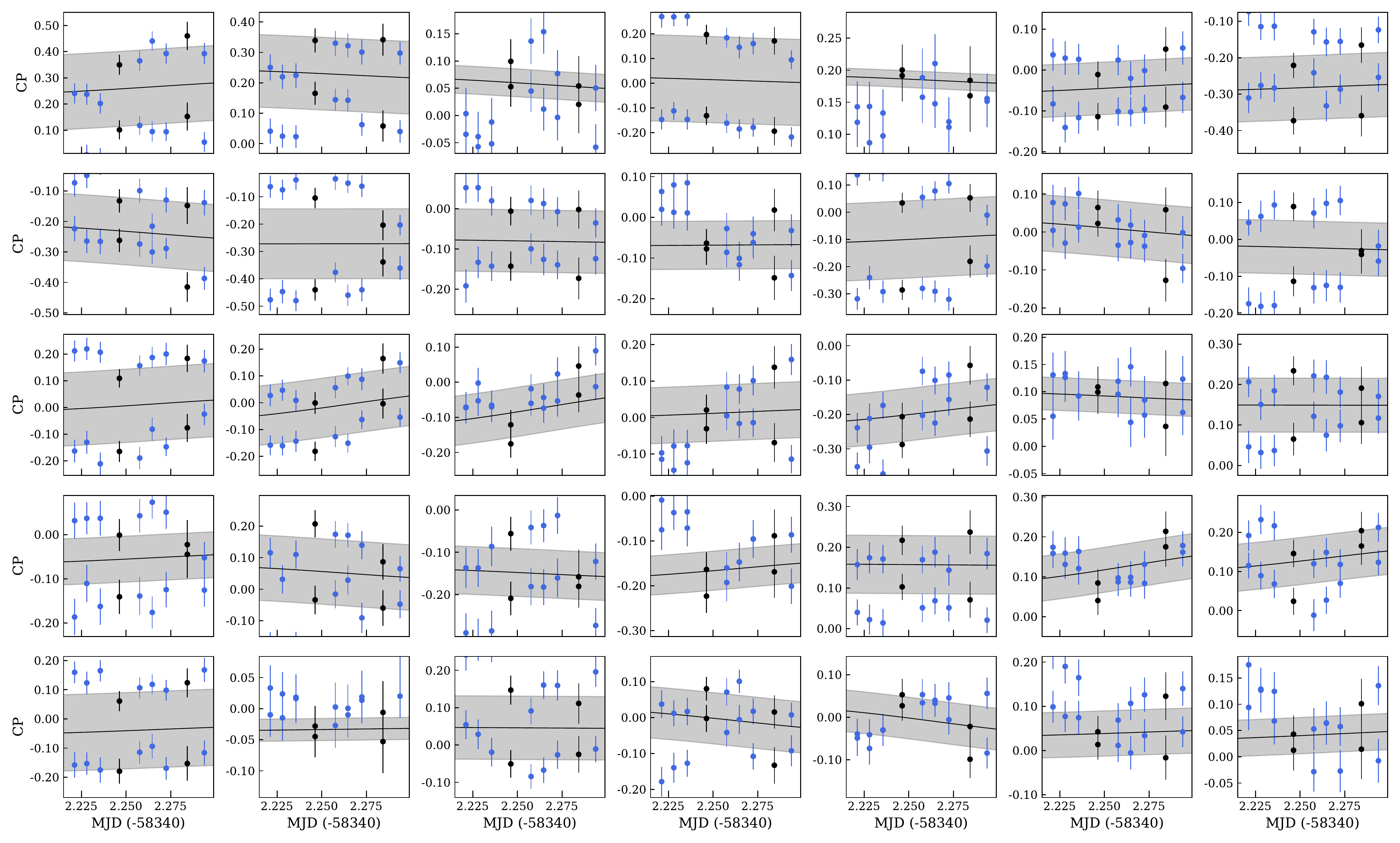}}
			\caption{Same as Fig.~\ref{figure__tf_v2} but for the closure phases.}
			\label{figure__tf_cp}
		\end{figure*}
		\FloatBarrier
		
		\section{Squared visibilities and closure phases for all observations}
		
		\begin{figure*}[!h]
			\resizebox{\hsize}{!}{\includegraphics{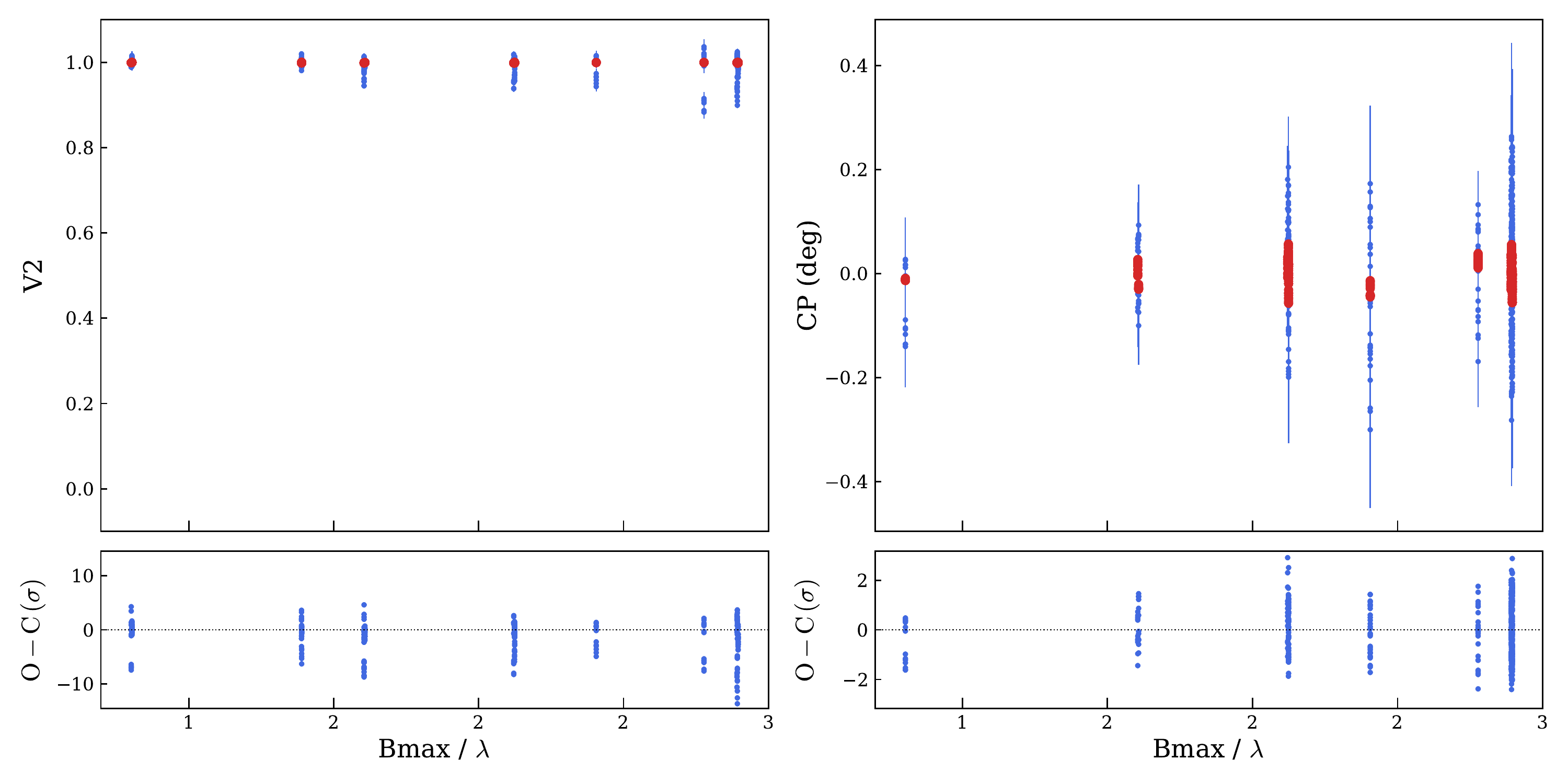}\includegraphics{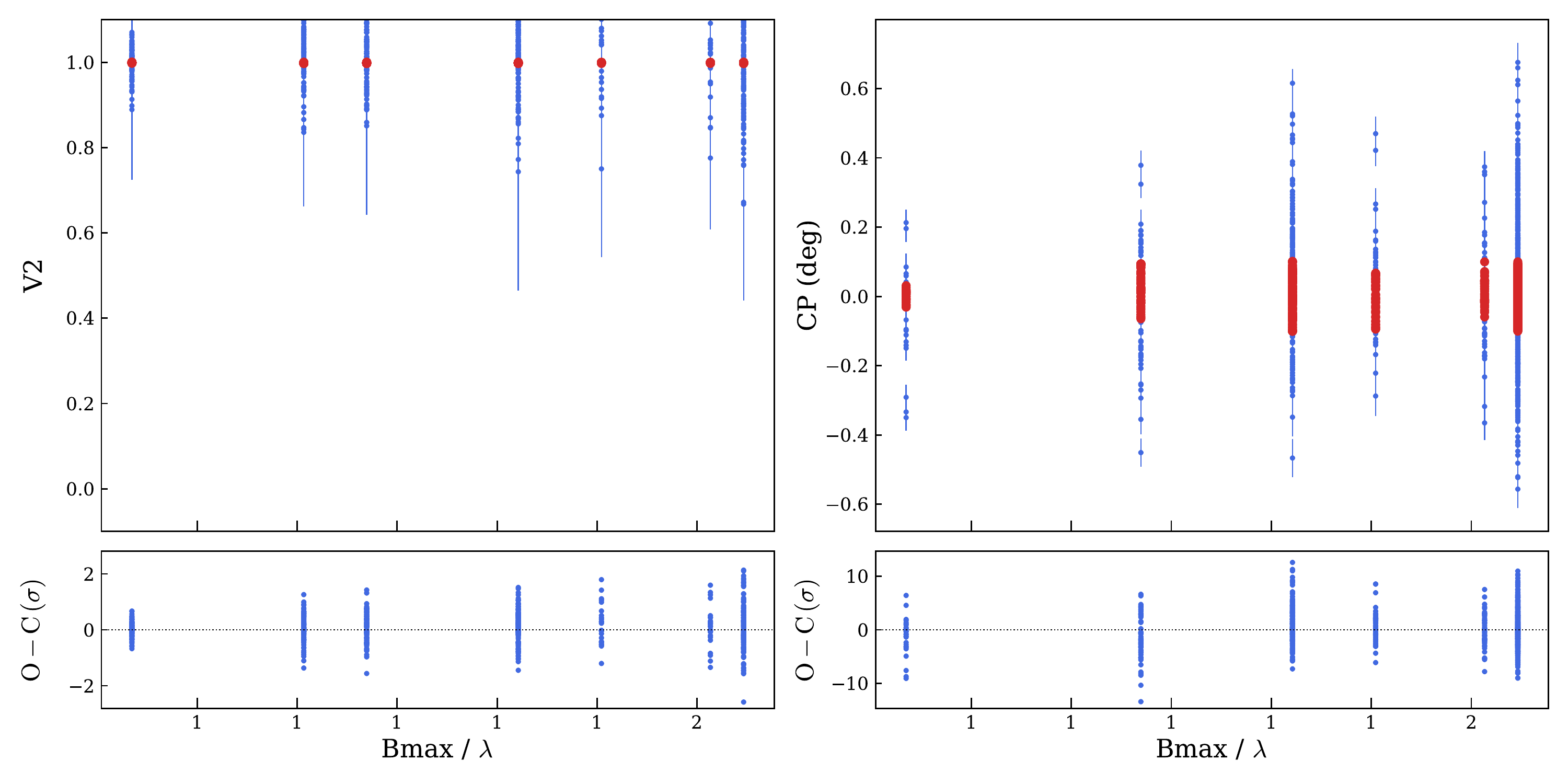}}
			\caption{Squared visibilities and closure phases for the SPHERE (left) and NACO (right) observations (blue). The red dots correspond to the best model found with \texttt{CANDID}.}
			\label{figure__sphere_naco_data}
		\end{figure*}
	\FloatBarrier
		\begin{figure*}[!h]
			\resizebox{\hsize}{!}{\includegraphics{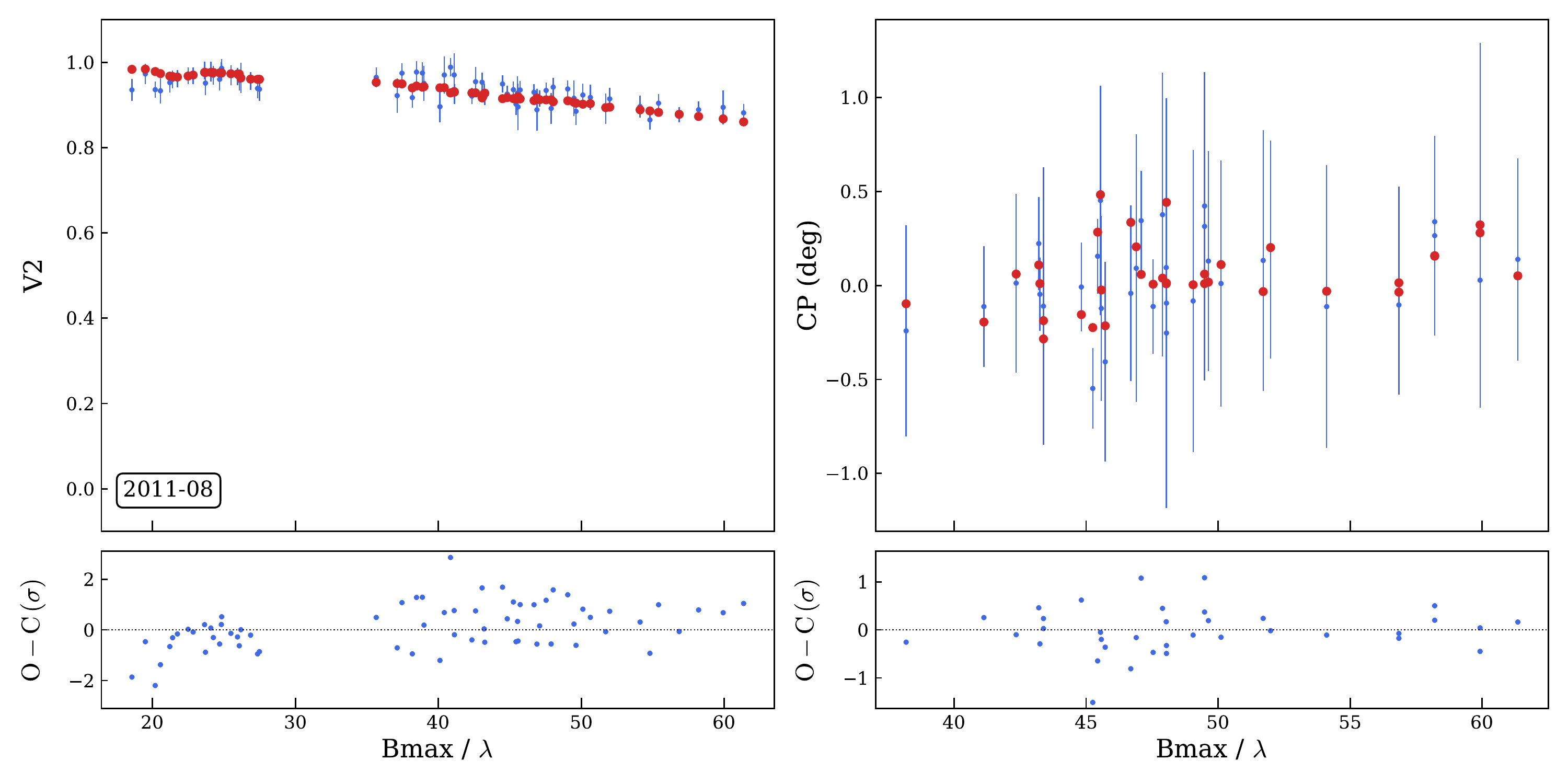}\includegraphics{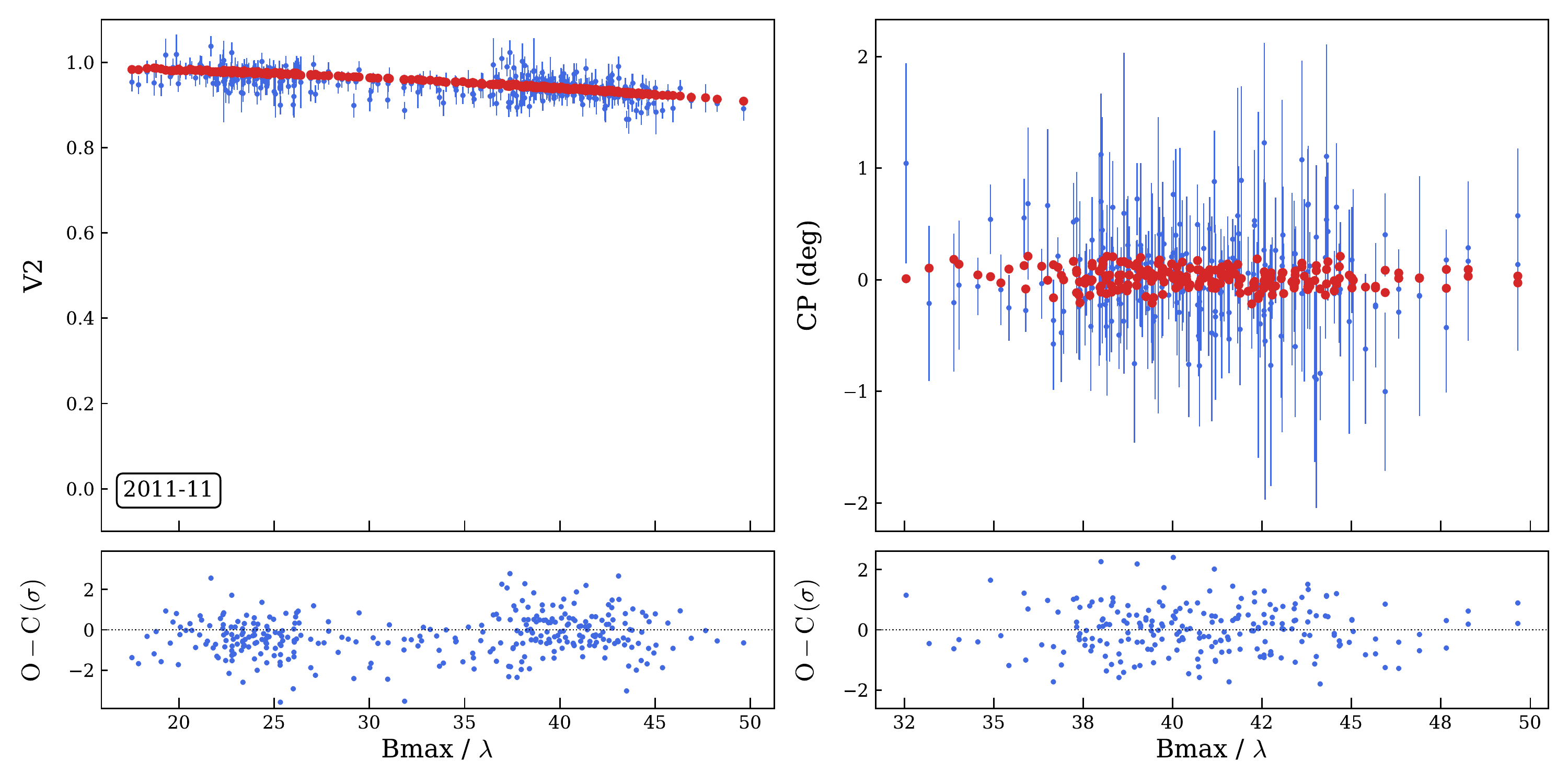}}
			\resizebox{\hsize}{!}{\includegraphics{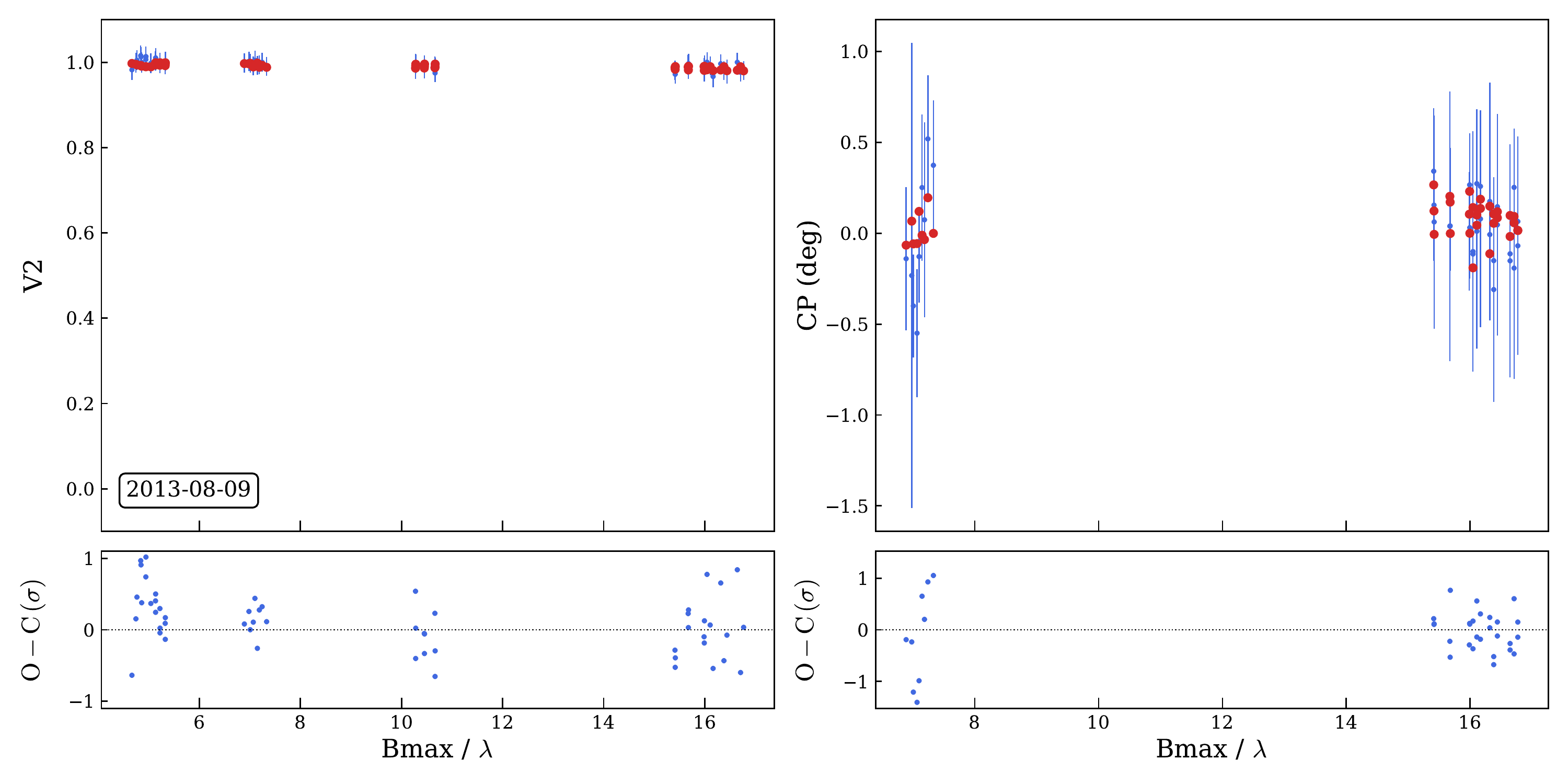}\includegraphics{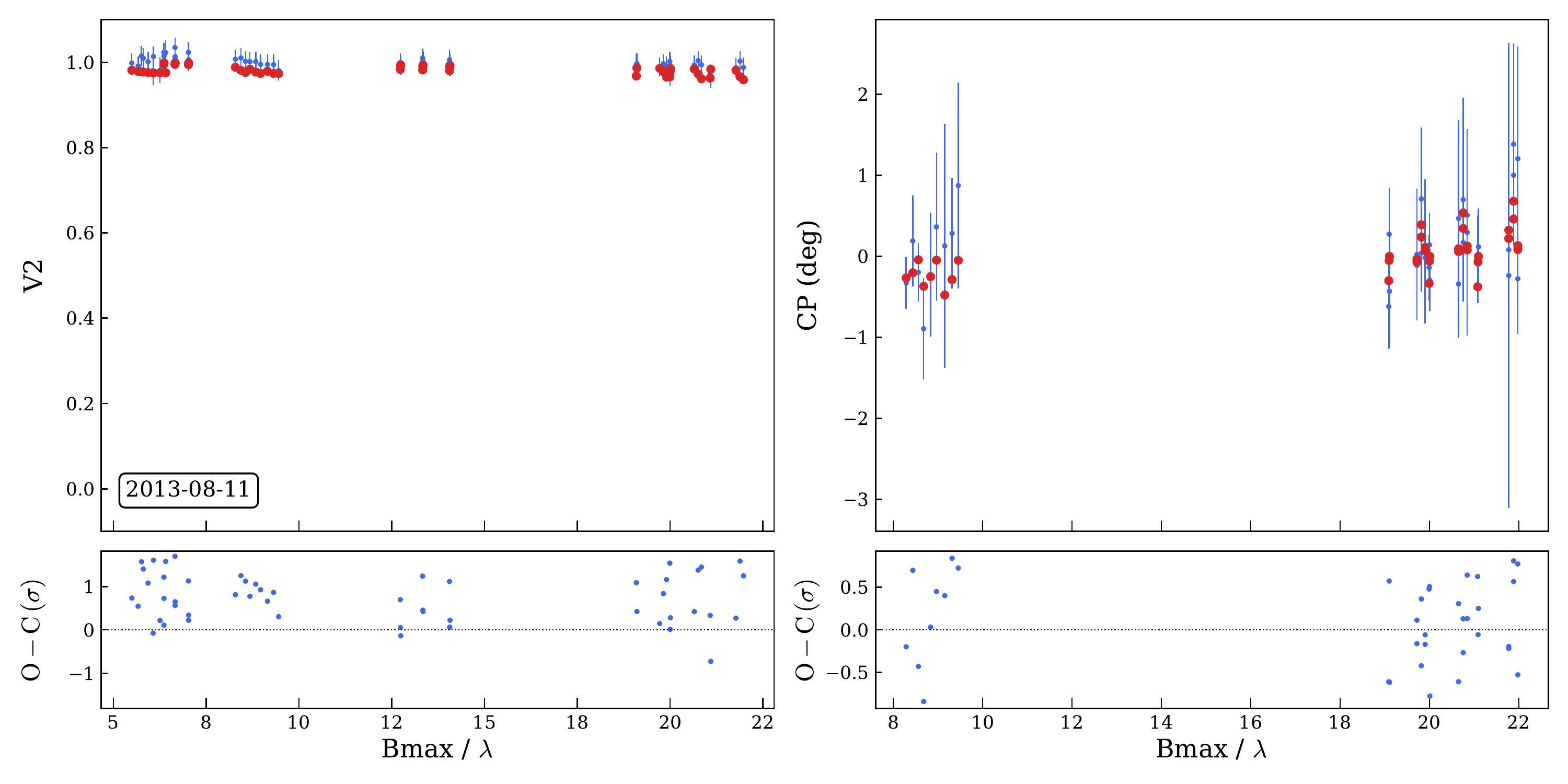}}
			\resizebox{\hsize}{!}{\includegraphics{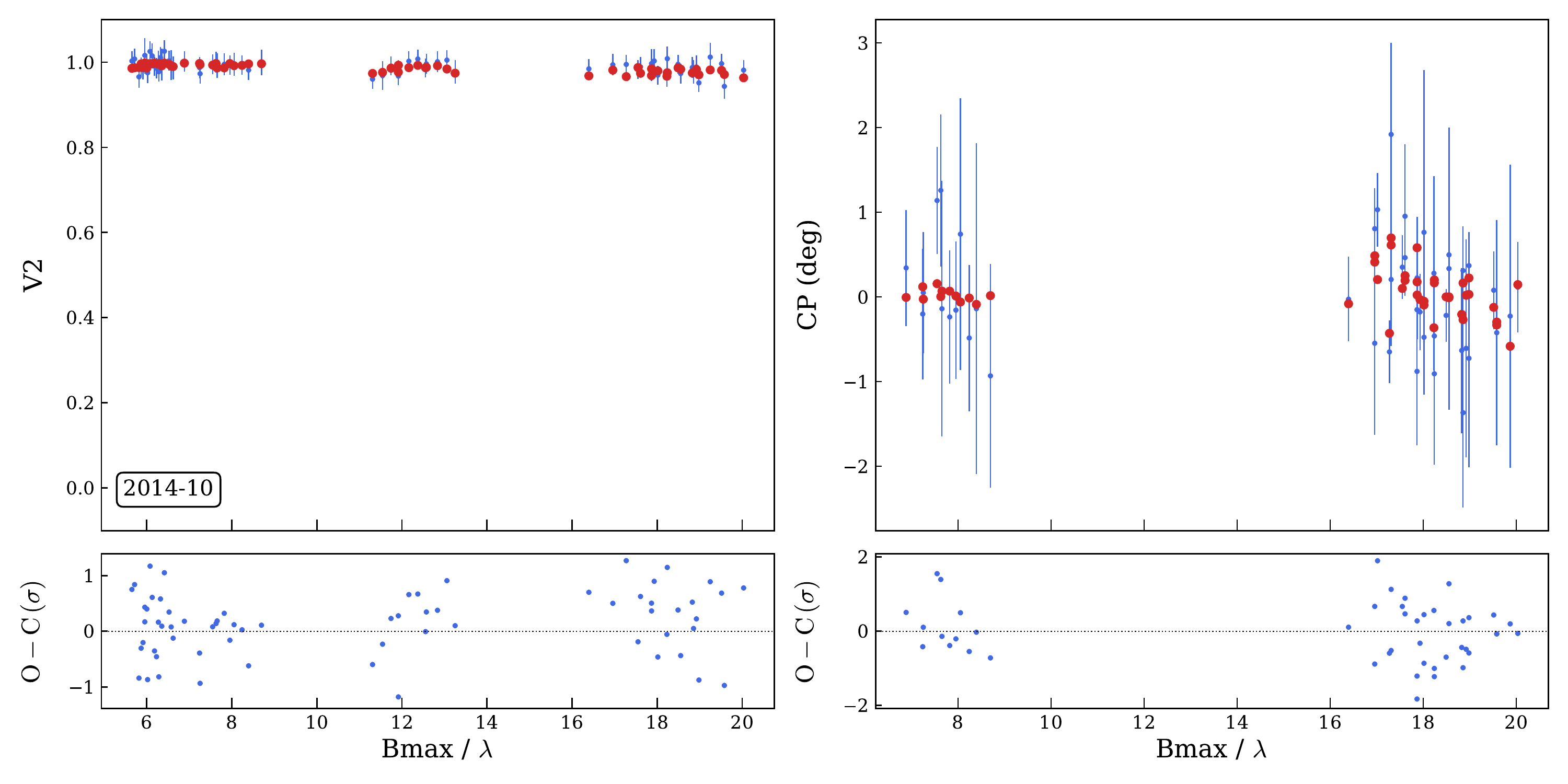}\includegraphics{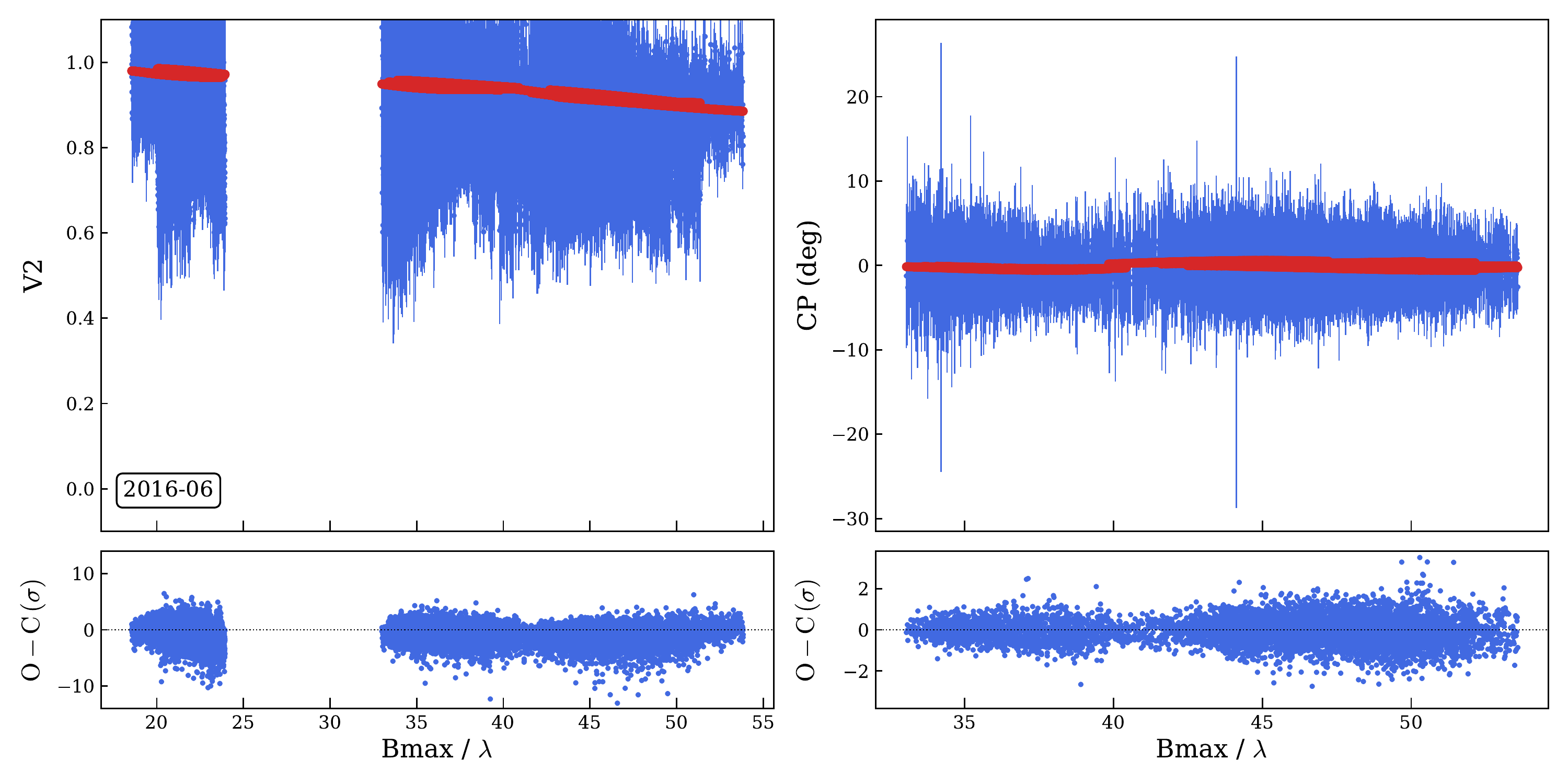}}
			\caption{Squared visibilities and closure phases for the PIONIER and GRAVITY observations (blue). The red dots correspond to the best model found with \texttt{CANDID}.}
			\label{figure__pionier_gravity_data}
		\end{figure*}
		\FloatBarrier
		
		\section{Detection maps for the NACO, PIONIER and GRAVITY observations}
		\label{appendix__detection_maps}
		
		\begin{figure*}[h]
			\resizebox{\hsize}{!}{\includegraphics{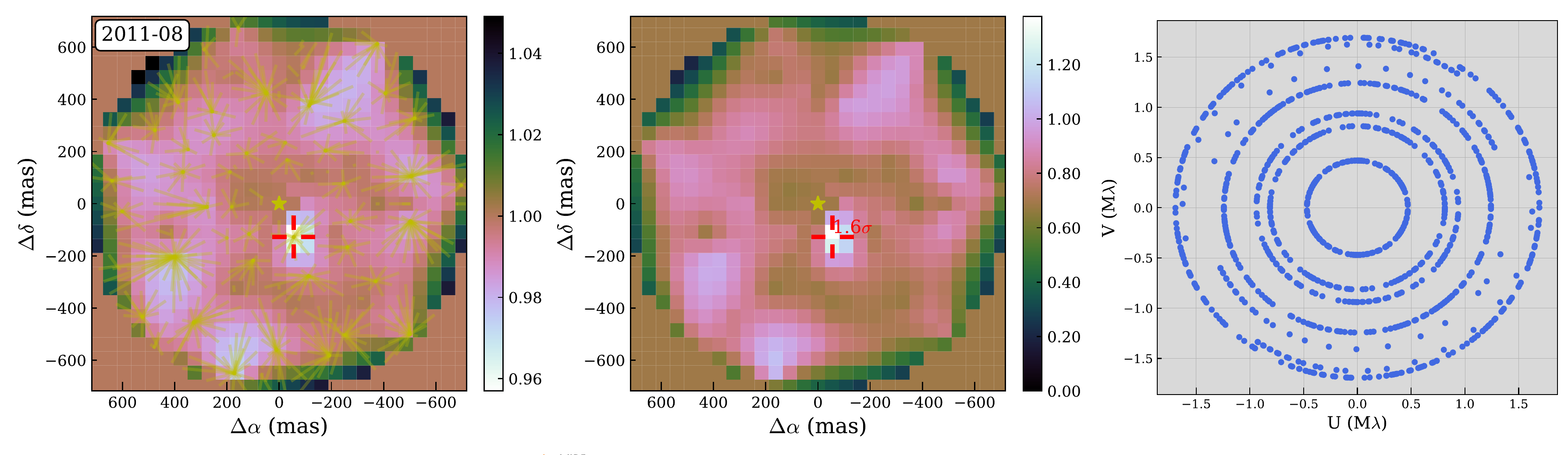}}
			\caption{$\chi^2_r$ map of the local minima for the NACO observations (left panel), detection level map (middle panel), and the $(u,v)$ plane coverage of the observations (right panel). The yellow lines represent the convergence from the starting points to the final fitted position. The maps were re-interpolated in a regular grid for clarity. The yellow star denotes AU~Mic, while the orange arrow in the middle indicates the disc orientation.}
			\label{figure_naco_detection_map}
		\end{figure*}
	    \FloatBarrier
	
		\begin{figure*}[h]
			\resizebox{\hsize}{!}{\includegraphics{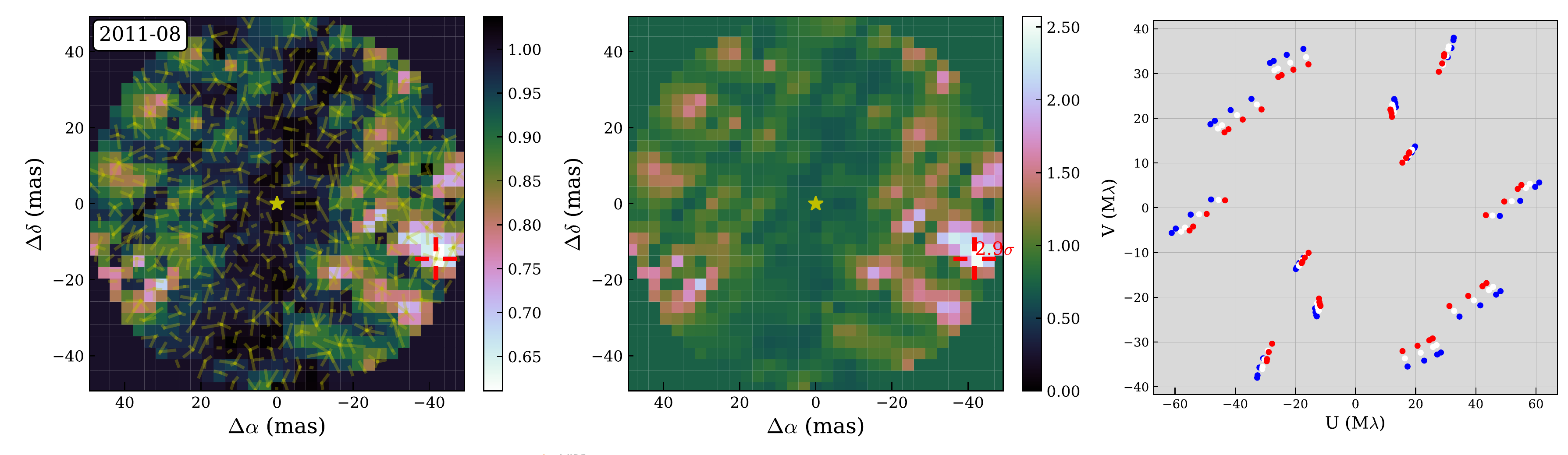}}
			\resizebox{\hsize}{!}{\includegraphics{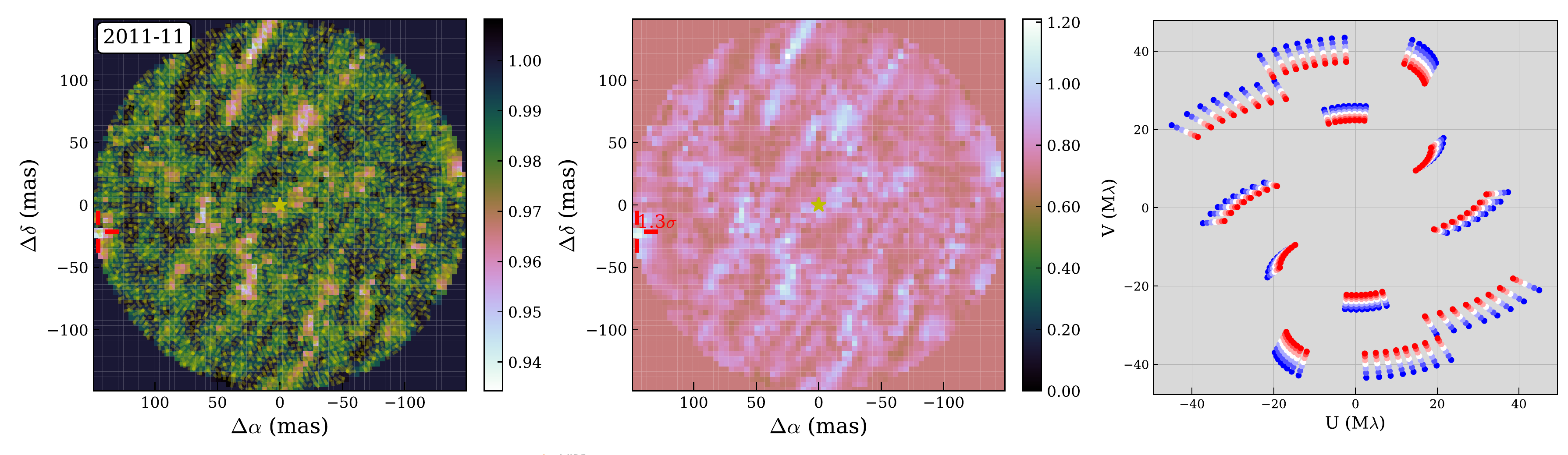}}
			\resizebox{\hsize}{!}{\includegraphics{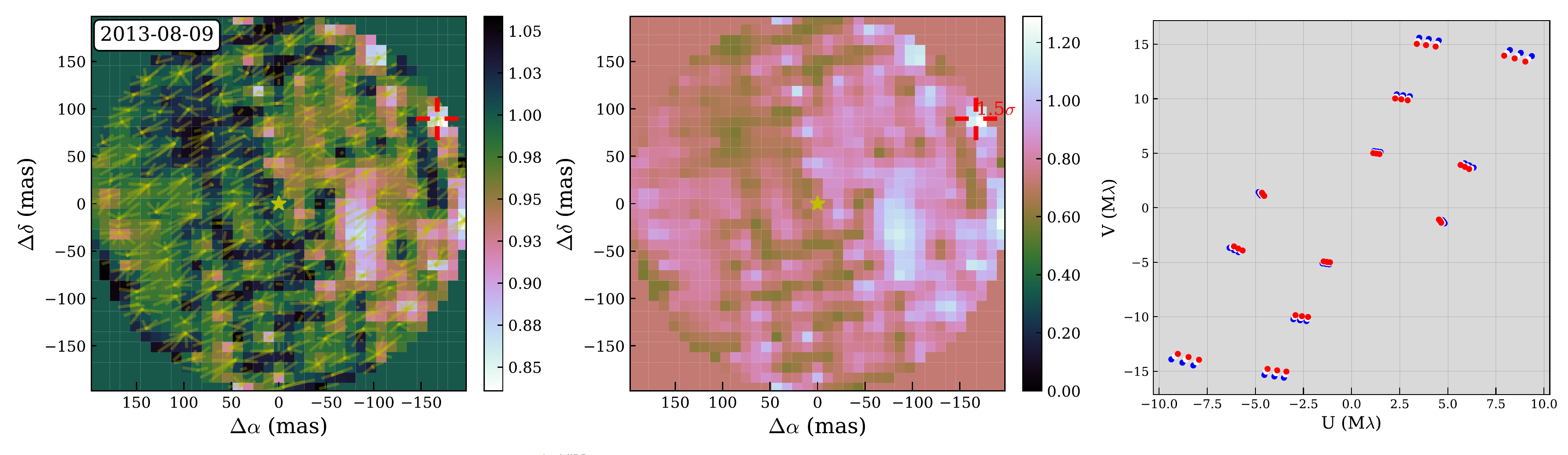}}
			\resizebox{\hsize}{!}{\includegraphics{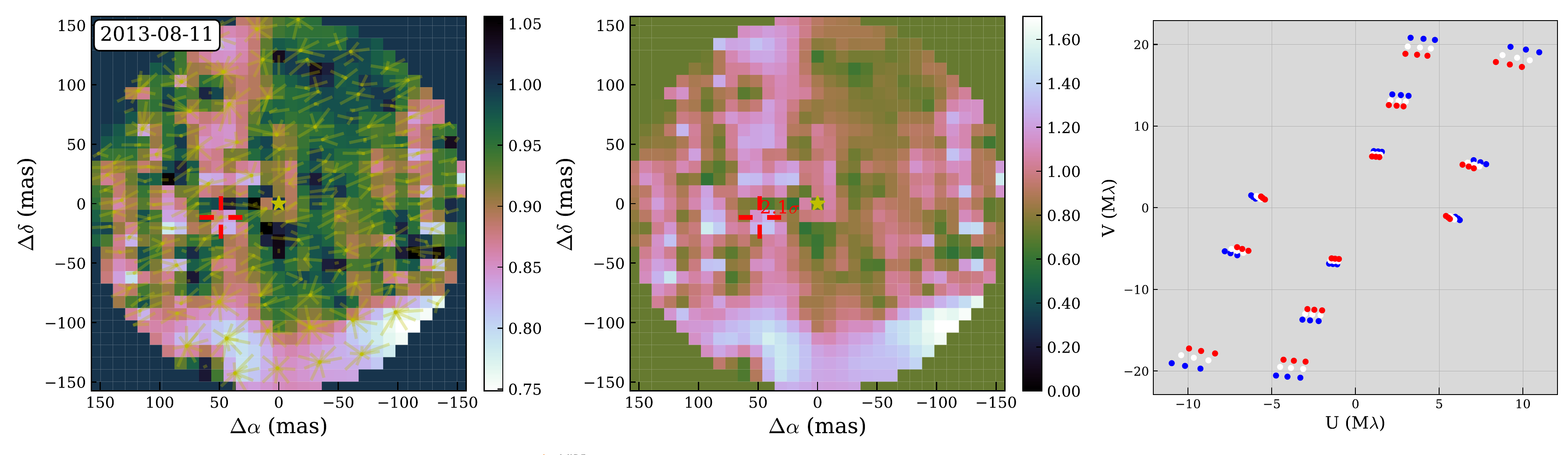}}
			\caption{Same as Fig.~\ref{figure_naco_detection_map} but for the PIONIER observations of 2011 and 2013.}
		\end{figure*}
		\FloatBarrier
		\begin{figure*}[h]
			\resizebox{\hsize}{!}{\includegraphics{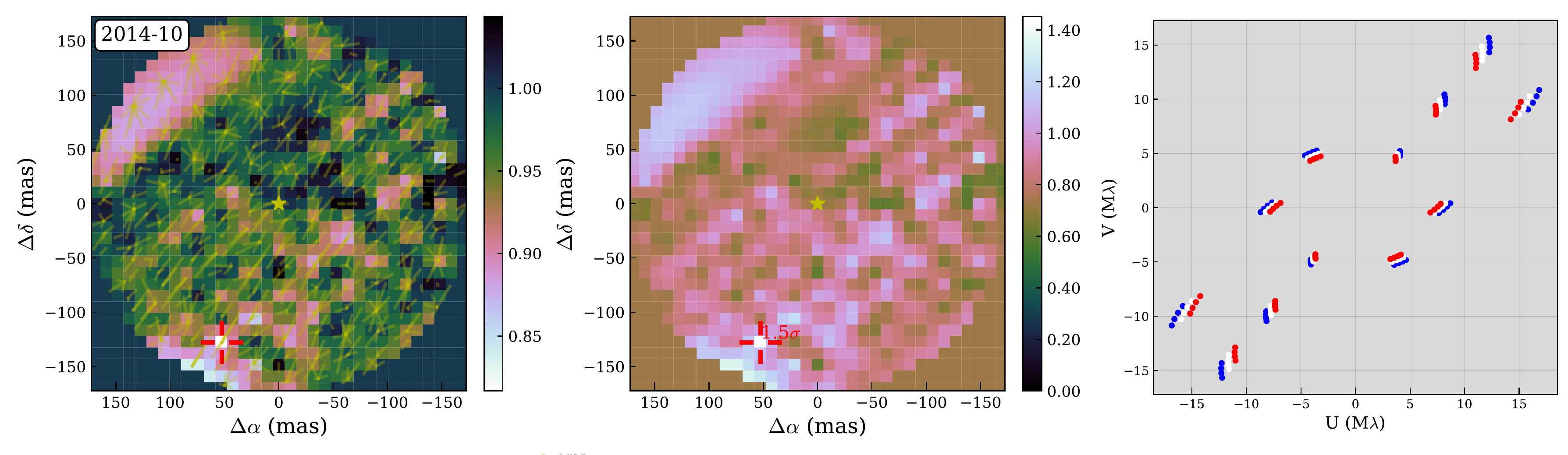}}
			\resizebox{\hsize}{!}{\includegraphics{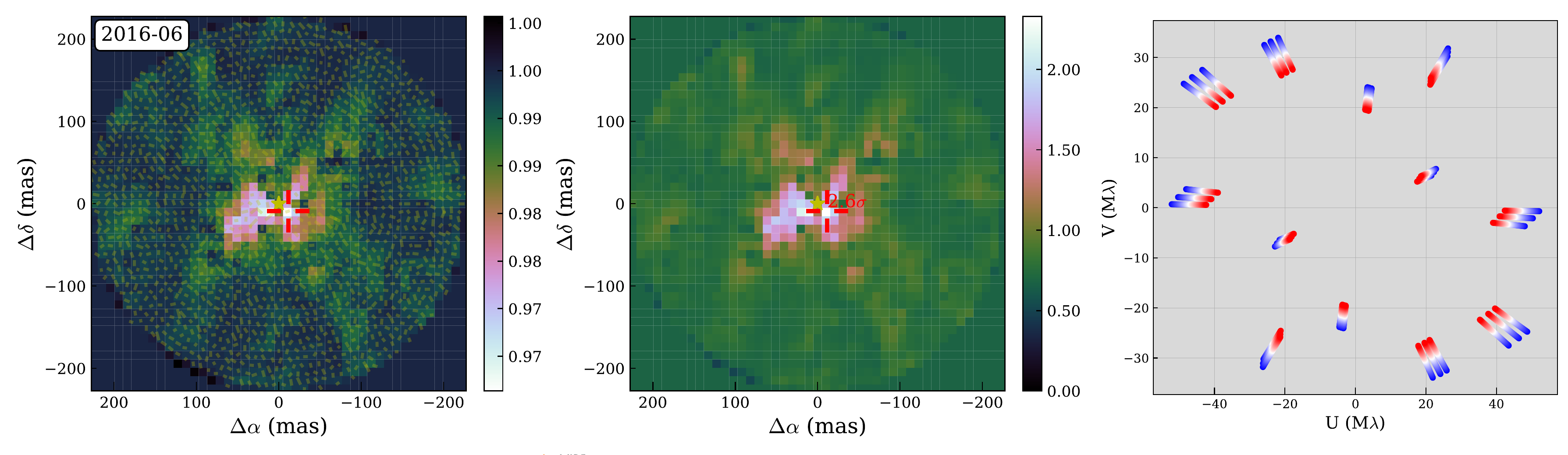}}
			\caption{Same as Fig.~\ref{figure_naco_detection_map} but for the 2014 PIONIER (top) and GRAVITY (bottom) observations.}
		\end{figure*}	
		\FloatBarrier
		
	\end{appendix}
	
\end{document}